\documentclass[prodmode,acmtomm,dvipsnames]{acmsmall} 

\makeatletter 
\newcommand\subparagraph{%
  \@startsection{subparagraph}{5}
  {\parindent}
  {3.25ex \@plus 1ex \@minus .2ex}
  {-1em}
  {\normalfont\normalsize\bfseries}}
\makeatother

\usepackage{epsfig}
\usepackage{amsmath}
\usepackage{amssymb}
\usepackage{graphics}
\usepackage{color}
\usepackage{xcolor}
\usepackage{multirow}
\usepackage{caption}
\usepackage{cite}
\usepackage{array}
\usepackage{url}
\usepackage{subfig}
\usepackage{bibspacing}
\usepackage{slashbox}
\usepackage{placeins}
\usepackage{titlesec}
\usepackage{helvet}
\usepackage{bm}
\usepackage[normalem]{ulem}

\usepackage[ruled]{algorithm2e}
 \newcommand{\ignore}[1]{}
\newboolean{showcomments} \setboolean{showcomments}{true}
\newcommand{\addref}[0]{\ifthenelse{\boolean{showcomments}} {
    \textcolor{green}{(Add ref(s))}}{}}

\titleformat{\subsubsection}[runin]
  {\normalfont\fontsize{9}{11}\sffamily\bfseries\slshape} 
  {\thesubsubsection}
  {1em}
  {}
  
  \titleformat{\paragraph}[runin]
  {\normalfont\fontsize{9}{11}\sffamily\bfseries\slshape} 
  {\thesubsubsection}
  {1em}
  {}

\SetAlFnt{\small}
\SetAlCapFnt{\small}
\SetAlCapNameFnt{\small}
\SetAlCapHSkip{0pt}
\IncMargin{-\parindent}

\begin{document}

\markboth{V. Gokhale et al.}{Congestion Control for Network-Aware Telehaptic Communication}

\title{Congestion Control for Network-Aware Telehaptic Communication}
\author{Vineet Gokhale
\affil{Indian Institute of Technology Bombay}
Jayakrishnan Nair
\affil{Indian Institute of Technology Bombay}
Subhasis Chaudhuri
\affil{Indian Institute of Technology Bombay}
}

\begin{abstract}
  Telehaptic applications involve delay-sensitive multimedia
  communication between remote locations with distinct Quality of
  Service (QoS) requirements for different media components. These QoS
  constraints pose a variety of challenges, especially when the
  communication occurs over a shared network, with unknown and
  time-varying cross-traffic. In this work, we propose a transport
  layer congestion control protocol for telehaptic applications
  operating over shared networks, termed as \textit{dynamic
    packetization module} (DPM). DPM is a lossless, network-aware
  protocol which tunes the telehaptic packetization rate based on the
  level of congestion in the network. To monitor the network
  congestion, we devise a novel \textit{network feedback module},
  which communicates the end-to-end delays encountered by the
  telehaptic packets to the respective transmitters with negligible
  overhead. Via extensive simulations, we show that DPM meets the QoS
  requirements of telehaptic applications over a wide range of network
  cross-traffic conditions. We also report qualitative results of a
  real-time telepottery experiment with several human subjects, which
  reveal that DPM preserves the quality of telehaptic activity even
  under heavily congested network scenarios. Finally, we compare the
  performance of DPM with several previously proposed telehaptic
  communication protocols and demonstrate that DPM outperforms these
  protocols.
\end{abstract}


\begin{bottomstuff}
Authors' address: Department of Electrical Engineering, Indian Institute of Technology Bombay, Mumbai, India 400076;\\
email: \{vineet, jayakrishnan.nair, sc\}@ee.iitb.ac.in
\end{bottomstuff}

\maketitle

\section{Introduction}

\label{sec:introduction}
Telehaptic applications, such as telesurgery \cite{ref:anderson},
involve long distance transfer of haptic-audio-visual information
between distantly located users. The performance of a telehaptic
activity is governed by a set of Quality of Service (QoS) parameters,
specific to each type of media. According to
\cite{ref:audioqosmiras,ref:qos,ref:interactiveVideo},
the QoS one-way delay and jitter specifications for multimedia are,
respectively, as follows: interactive video - 400 ms and 30 ms, audio
- 150 ms and 30 ms, haptic - 30 ms and 10 ms.
Non-conformance to the above constraints leads to degraded human
perception, and can potentially compromise the quality of the
telehaptic activity \cite{ref:hapticdelay}. In particular, a haptic
QoS violation results in destabilizing the haptic global control loop
\cite{ref:unstableferrell,ref:anderson}, and a deteriorated perception
of haptic objects. Hence, multimedia data reception and display within
the prescribed QoS deadlines plays a pivotal role in determining the
stability and the overall performance of a telehaptic task.

In a shared network, like the Internet, a telehaptic source shares the
network resources with other concurrent traffic streams. As a result,
the intensity of the cross-traffic encountered by a telehaptic stream
on a shared network is both unknown as well as time-varying. In such a
scenario, the transmission of telehaptic data in a
\emph{network-oblivious} manner can be highly sub-optimal. In
particular, at times when the network is severely congested, a
network-oblivious telehaptic stream may suffer large delays and
frequent packet losses, leading to QoS violations. Note that this is
all the more likely in resource constrained networks, such as wireless
ad-hoc networks. On the other hand, at times when the network is
lightly loaded, it may be feasible to transmit telehaptic data at its
peak rate. The above discussion motivates the need for a
\emph{network-aware} telehaptic transmission scheme. In this paper, we
propose such a scheme, which monitors network conditions in real-time
and adapts the telehaptic data rate accordingly to achieve congestion
control in a lossless manner.

\ignore{The amount of network resource available for the telehaptic
  application reduces with the increasing number of users, and the
  quantity of data injected into the network. Disaster management
  applications, that involve deploying ad-hoc networks, face severe
  limitations on the communication resources.  The scarcity of the
  resources makes the application more susceptible to QoS
  violation. Additionally, the variable data rate nature of the
  internet traffic creates network irregularities, like jitter and
  packet losses.  The resulting large delay, jitter and packet losses
  can potentially cause damage to the ongoing telehaptic application.
  In order to allow the network to accommodate other users, while
  satisfying the telehaptic QoS requirements, the telehaptic sources
  should operate in a network-aware manner so that the telehaptic data
  rate is tuned to suit the currently available network resources.
  Hence, a telehaptic application equipped with a network adaptive
  scheme is more robust on a shared network. Additionally, if the rate
  adaptation operates without introducing any loss to the telehaptic
  data, then the telehaptic application provides a high fidelity
  display of media at the receivers.  An early detection of the
  unfavorable network changes plays a key role in telehaptic delay
  control.  In this article, we seek to design a network-aware
  communication protocol for telehaptic data rate adaptation on a
  shared network.}

\subsection{Contributions of the Article}
\label{subsec:contrib}

  In this article, we focus on point-to-point telehaptic
  communication over a shared network. Specifically, we propose a
  network-aware protocol for multiplexing and transmission of
  haptic/audio/video data between two telehaptic nodes connected via a
  shared network. The protocol monitors network congestion in real
  time and (losslessly) adapts the transmission rate on the forward
  and the backward channels to maintain QoS compliance.

  The proposed protocol receives haptic/audio/video frames from the
  respective capture devices at each node and delivers these frames to
  the corresponding display devices at the other end. By design, our
  protocol is robust to the type and resolution of the media devices,
  as well as the audio/video encoding standard employed. Thus, our
  protocol may be viewed as a transport layer congestion control
  solution for point-to-point telehaptic communication, akin to the
  Transmission Control Protocol (TCP) for elastic internet traffic.

  The proposed protocol has two main components:

  \begin{enumerate}
   
   \item \emph{Network feedback module:} The network feedback module (see
  Section~\ref{subsubsec:delaynotif}) is a novel mechanism for real time monitoring of
  end-to-end delays on the network. It exploits the bidirectional
  nature of telehaptic traffic to convey delays on each channel to the
  corresponding transmitter. Specifically, the end-to-end delays as
  measured by a receiving node are piggybacked on telehaptic data
  packets on the reverse channel (see Figure~\ref{fig:delaynotif}). This provides
  real-time feedback of network state to the transmitting node with
  negligible overhead (3 bytes per packet). The proposed network
  feedback module can also potentially be utilized for other
  (bidirectional) media streaming applications like video
  conferencing.
  
  \item \emph{Dynamic packetization module:} The dynamic packetization
  module (DPM) (see Section~\ref{subsubsec:networkbased}) is a lossless mechanism for
  telehaptic data rate adaptation, based on the delay feedback from
  the network feedback module. DPM is motivated by the following
  observation: Under telehaptic data transmission at the default
  packetization rate of 1000 packets/sec, the overhead due to packet
  headers from various layers accounts for almost half the total
  telehaptic traffic. Thus, when the network is congested, DPM
  dynamically merges successive telehaptic fragments into a single
  packet, thereby lowering the overall transmission rate to match the
  available network capacity. Naturally, this transmission rate
  reduction is achieved at the expense of additional packetization
  delay at the transmitter.
  
  \end{enumerate}

  We evaluate the proposed telehaptic transmission scheme via
  extensive simulations as well as human subjective tests through a
  real-time telepottery experiment (see Section~\ref{sec:results}). Our
  simulations reveal that DPM meets the telehaptic QoS specifications
  even under extremely congested network settings. Our subjective
  tests confirm that DPM provides a seamless telehaptic user
  experience in a congested network. We also compare DPM with other
  recently proposed telehaptic communication protocols, and demonstrate
  that DPM outperforms these protocols with respect to QoS compliance.

  Finally, to evaluate the performance of the proposed scheme
  analytically, we derive bounds for the maximum haptic/audio/video
  delay on a network with a single bottleneck link, assuming constant
  bit-rate (CBR) cross-traffic (see Appendix~\ref{sec:upperbound}
  and~\ref{sec:AVbound}). These delay characterizations are useful
  in identifying network settings where QoS-compliant telehaptic
  communication is feasible.

\subsection{Related Work}
\label{subsec:relatedwork}
There have been several attempts to address the problem of large
telehaptic bandwidth requirement. The standard input and output update
rate of the haptic signal is 1 kHz. In order to reduce the
packetization delay encountered by the haptic samples, the
conventional approach follows fixed haptic packetization at 1 kHz (1
packet per sample) for transmission over the network.  This approach
is highly bandwidth demanding, and is not friendly to other network
users. To counter this issue, the works in
\cite{ref:novel,ref:telepresence,ref:psychophysics,ref:perception,ref:weberlaw,ref:exploring}
explored \textit{adaptive sampling} which exploits the perceptual
limitation of the human haptic system to achieve lossy haptic signal
compression.  A \textit{just noticeable difference} (JND) metric
adaptively marks the haptic samples that are not perceivable by the
human users. The communication system refrains from transmitting such
samples, thereby reducing the telehaptic data rate. The missing haptic
samples are then reconstructed at the receiver using standard
extrapolation techniques listed in \cite{ref:hoip}. However, critical
operations, like telesurgery, necessitate accurate replication of the
surgeon's hand movements. In such scenarios, a minor loss of precision
due to adaptive sampling could result in potentially irreparable
damage. Also, a teleoperator, such as a robotic device, could
practically sense all haptic samples; in such cases, adaptive sampling
discards perceptually significant samples. Another networking-related
issue with adaptive sampling is the following. The
\emph{instantaneous} source rate of adaptive sampling depends purely
on the speed of haptic interaction, and can at times far exceed the
\emph{average} source rate. As a result, provisioning the network for
the average source rate can lead to serious QoS violations; this is
demonstrated in Section~\ref{sec:comparison}. In other words, adaptive
sampling does not provide any real economies with respect to network
bandwidth requirement -- one needs to provision network capacity for
the \emph{peak} telehaptic data rate in order to avoid QoS violations.

\ignore{Therefore, even though the long run average source rate is
  very low the instantaneous rates can be as high as the peak rate.
  Hence, from a networking standpoint, adaptive sampling is not an
  effective means of compressing haptic data as we demonstrate in our
  results.  To summarize, the lossy compression achieved by adaptive
  sampling is not always a suitable means of reducing the data
  rate. Moreover, note that the data rate reduction due to adaptive
  sampling is network-oblivious.  In this work, we propose a lossless,
  network-aware scheme for telehaptic communication.
}

Several application layer protocols have been specifically designed
for telehaptic communication.  ALPHAN: Application Layer Protocol for
HAptic Networking, proposed in \cite{ref:alphan}, implements haptic
and graphic data communication at the packetization rate of 1
kHz. AdMux: Adaptive Multiplexer \cite{ref:admux} proposes a
statistical multiplexing scheme for scheduling haptic-audio-video
packet transmission based on the QoS requirements and changing network
behavior.  Haptics over Internet Protocol (HoIP) for point-to-point
communication, proposed in \cite{ref:vineetncc}, addresses media
multiplexing and telehaptic communication involving haptics, audio and
video data. The above mentioned protocols carry out telehaptic
transmission at the peak rate, and hence do not address the problem of
congestion control.  In \cite{ref:cizmeci}, the authors consider
visual-haptic multiplexing over constant-bitrate (CBR) communication
links, employing adaptive sampling for haptic signal compression.
However, the drawbacks of adaptive sampling mentioned previously apply
here; see Section~\ref{sec:comparison} for a demonstration.

\ignore{Adaptive sampling minimizes the telehaptic bandwidth
  requirement to a great extent, but at the expense of reduced quality
  of haptic signal reconstruction, as mentioned earlier.  Moreover,
  when the network is in a condition to support 1 kHz packetization
  rate, the adaptive sampling scheme unnecessarily discards a
  significant amount of haptic samples, thereby losing an opportunity
  for a more precise teleoperation. The aforementioned haptic
  protocols remain oblivious to network changes, and therefore are not
  suited for real-world networks.}

The work in
\cite{ref:packetizationintfujimoto} explores the possibility of
merging multiple haptic samples in a packet to reduce the telehaptic
data rate. In contrast with the scheme proposed in this paper, the
scheme in \cite{ref:packetizationintfujimoto} always combines a fixed
number of haptic samples, irrespective of the network conditions. Note
that this implies unnecessary packetization delay even when the
network is uncongested. Moreover, the authors showed in a particular
setting that a packetization interval of 8 ms results in a
satisfactory user performance. On the contrary, we demonstrate (see
Figure \ref{fig:kplot}) that the packetization intervals greater than
4 ms result primarily in increasing end-to-end delays, without any
substantial reduction in the telehaptic data rate.

Note that the above mentioned proposals are all
\emph{network-oblivious}, i.e., they do not adapt the telehaptic
transmission rate based on network conditions. The literature provides
a few works that have considered \emph{network-aware} telehaptic rate
adaptation. We discuss these next.

In \cite{ref:netadaplee}, the authors propose a network adaptation
scheme for merging haptic samples based on packet losses arising out
of congestion. Such a scheme is reactive to network congestion, in the
sense that data rate reduction is activated only after detecting
persistent packet losses. Clearly, such a loss-based congestion
control mechanism is not suitable for highly delay-sensitive
telehaptic applications.
We note that \cite{ref:netadaplee} does not provide much detail about
the rate adaptation mechanism itself; also, the effects of this rate
adaptation on other concurrent network flows are not analyzed.

The authors in \cite{ref:wirzetp} propose the first delay-sensitive
haptic communication protocol named Efficient Transport Protocol
(ETP). ETP detects congestion based on round-trip-time (RTT)
measurements. Once congestion is detected, ETP reduces the telehaptic
data rate by increasing the interpacket gap, i.e., by
\emph{downsampling} the haptic signal. In contrast, the protocol
proposed in this paper preserves the fidelity of the haptic signal,
adapting instead the packetization rate based on the congestion level
in the network.

The paper most closely related to ours is \cite{ref:nafcahkokkonis},
which proposes NAFCAH: Network Adaptive Flow Control Algorithm for
Haptic data. Like DPM, NAFCAH adapts the number of haptic samples to
be merged into a packet on the forward channel based on network
conditions. However, there are two crucial differences between NAFCAH
and DPM. First, when congestion is detected, NAFCAH decreases its
transmission rate in stages. In contrast, DPM responds to congestion
with an aggressive rate reduction, which enables network buffers to
get flushed quickly, minimizing the possibility of QoS violations. Second, NAFCAH
monitors congestion based on RTT measurements. However, under
asymmetric network conditions, RTT may not provide an accurate
estimate of the (one-way) delay on the forward channel. In contrast,
DPM estimates the delay on the forward and backward channels
separately. The performance implications of the above differences are
demonstrated in Section~\ref{sec:comparison}.

The authors of \cite{ref:vineetopportunistic} propose a network-aware
opportunistic adaptive haptic sampling mechanism, wherein the adaptive sampling
threshold is varied based on the congestion level in the network. Note that
the limitations of adaptive sampling discussed earlier apply to this
work as well.

\ignore{The rate adaptation is also based on RTT, measured by
  regularly transmitting probing packets which consume additional
  network bandwidth. As we show in our results, such RTT-based
  adaptation results in a delayed congestion detection leading to
  serious QoS violations especially in an asymmetric network
  setting. Furthermore, unlike our approach, NAFCAH adopts multistep
  rate reduction, thereby exhibiting a sluggish response to
  congestion.  Moreover, this approach performs packetization up to 8
  haptic samples in a packet. As already mentioned, this leads to
  unnecessary packet delays without any significant data rate
  reduction.}

Finally, we contrast our work with the Real-time transport protocol
(RTP) \cite{ref:rtp}, which is the most commonly used protocol for
audio/video streaming and has also been recommended by some
researchers for telehaptic communication; see for example
\cite{ref:steinbachRTP}. RTP uses report-based notification for
monitoring the network conditions at regular intervals of time.  The
multimedia receiving agent sends RTP Control Protocol (RTCP) receiver
reports to the transmitters, for QoS monitoring, once in every 500 ms
\cite{ref:adaptive}. However, as we demonstrate in
Section~\ref{sec:comparison}, such sparse feedback is insufficient for
telehaptic applications, which are sensitive to network changes that
occur over a timescale of tens of milliseconds.

\subsection{Organization of the Article}
\label{subsec:organization}
The article is organized as follows. In Section
\ref{sec:telehapticdatacomm}, we describe the configuration of a
typical telehaptic environment, and explain in detail the design and
working of the proposed telehaptic communication framework. In Section
\ref{sec:exptdesign}, we discuss the setup for simulations and the
real-time telepottery experiment. Section \ref{sec:results} presents
the findings of the experiments, and in Section \ref{sec:conclusions},
we state our conclusions. Finally, in the appendix, we
  describe the DPM header structure, and characterize the maximum
  end-to-end haptic/audio/video delay under the proposed scheme in a
  simple (single bottleneck) network setting.

\section{Design of Telehaptic Communication Framework}
\label{sec:telehapticdatacomm}
In this section, we explain the standard telehaptic setting on a
shared network, and describe the techniques proposed in this article
for a lossless, network-aware, adaptive telehaptic data communication.

\subsection{Typical Telehaptic Environment}
\label{subsec:telehapticenvt}

\begin{figure}
\begin{center}
\includegraphics[height = 20mm, width = 60mm]{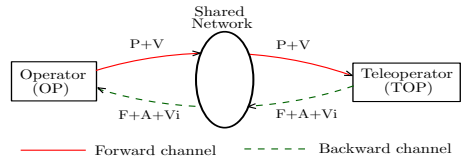}
\caption{A diagrammatic representation of master-slave based telehaptic setting showing the telehaptic data flow in a shared network.
Data flow notations: P - position, V - velocity, F - force, A - audio, Vi - video.}
\label{fig:telehaptic}
\end{center}
\end{figure}

We consider a typical point-to-point telehaptic application, like
telesurgery, running on a shared network as shown in
Figure~\ref{fig:telehaptic}.  The operator~(OP) acts as the master and
sends the current position and velocity commands to the
teleoperator~(TOP). In response, the TOP, acting as the slave,
transmits force information to the OP, in addition to auditory and
visual data.  The channels on which the operator and teleoperator
transmit telehaptic data are called forward and backward channels,
respectively. Note that the telehaptic traffic is bidirectional and
asymmetric in nature. Moreover, the forward and backward channel are
also asymmetric; in general, they may differ with respect to routing
paths, capacity, as well as cross-traffic. Our network feedback
module, described in Section~\ref{subsubsec:delaynotif}, estimates
congestion on the forward and the backward channels
separately. Finally, we remark that the particular master-slave setup
depicted in Figure~\ref{fig:telehaptic} is assumed only for
concreteness in exposition. Our proposed telehaptic communication
protocol works in any general point-to-point telehaptic application.

\ignore{Packetization at the
  typical haptic sampling rate of 1kHz leads to the telehaptic nodes
  transmitting a data packet in every milli-second. In reality, the
  telehaptic nodes share the network with several cross-traffic
  sources.  Throughout this article, we will discuss the working of
  the proposed techniques using the example of a typical master-slave
  based telehaptic activity on a shared network.}

\subsection{Media Multiplexing Framework}
\label{subsec:mediamultiplex}

In this section, we describe our media multiplexing
framework. Multiplexing the media frames appropriately from the
different capturing devices and forwarding them to the transmitter is
a critical task in any network based real-time interactive
application, since it directly influences the QoS adherence of the
respective media. The authors in \cite{ref:cizmeci} rightly explain
the importance of splitting a large video frame into smaller parts for
transmission. Naturally, if a large video frame is transmitted in a
single packet, it would clog the network for a long time, thereby
delaying the subsequent haptic/audio samples substantially. The media
multiplexing framework proposed here is an adaptation of that in
\cite{ref:cizmeci,ref:vineetncc}.

Our media multiplexer works in synchronization with the sampling of the
haptic signal, which we assume occurs at the default rate of 1
kHz. Each time a haptic sample is generated, our multiplexer generates
a telehaptic fragment of size $p,$ which contains the latest haptic
sample, as well as audio/video data as explained below.\footnote{If
  there is no audio/video data, as is the case in the communication
  from the OP to the TOP (see Fig.~\ref{fig:telehaptic}), then each
  telehaptic fragment is composed of a single haptic sample.}

  Let $f_a$ and $f_v$ denote the peak frame generation rate
  (in frames per second) for audio and video, respectively. Let $s_a$
  and $s_v$ denote the maximum size (in bytes) of an audio and video
  frame, respectively.\footnote{$f_a,$ $f_v,$ $s_a,$ and
      $s_v$ depend on the audio/video encoding standards employed. It
      is important to note that the proposed protocol, which operates
      at the transport layer, is robust to the encoding standards
      used.} The (peak) telehaptic payload generation rate, denoted
  by $D$ (in kbps), is expressed in terms of the individual media
  parameters as
\begin{equation}
 D = (f_h \cdot s_h + f_a \cdot s_a + f_v \cdot s_v)\cdot(8/1000).
 \label{equ:payloadrate}
\end{equation}
Here, $f_h$ denotes the haptic sampling rate (assumed to be 1 kHz),
and $s_h$ denotes the size of a haptic sample.\footnote{Throughout
  this article, we use the terms sample and frame interchangeably.}

In order to maintain equilibrium between the payload generation and
the multiplexing, the size $p$ (in bytes) of the telehaptic fragment
multiplexed per milli-second is given by
\begin{equation}
 p = \frac{D}{8} (1 ms).
 \label{equ:payloadsize}
\end{equation}
Due to the mandatory haptic sample in each telehaptic fragment, the
size of audio/video data in a fragment is given by
\begin{equation}
 s_m = p-s_h .
 \label{equ:fragsize}
\end{equation}
Since audio has a stricter QoS constraint than video, our multiplexer
gives audio data priority over video data. That is, in each telehaptic
fragment, the multiplexer packs $s_m$ bytes of audio/video data (not
previously multiplexed), giving strict priority to audio over
video. It can be shown that the proposed hierarchical
  priority-based multiplexing mechanism leads to substantially lower
  audio/video jitter compared to the first-come-first-served
  multiplexing mechanism proposed in \cite{ref:vineetncc}.

\subsection{Network Feedback Module}
\label{subsubsec:delaynotif} 

The network feedback module performs two functions: i) it monitors the
delays on the forward and backward channels separately through \textit{in-header delay notification} mechanism, and ii) based on the received
piggybacked delays it generates
triggers for the respective transmitters to adapt their data rates. We
explain these functions in the following.

\subsubsection{In-header delay notification}
\begin{figure}[!h]
\begin{minipage}[b]{0.45\linewidth}
\centering
    \includegraphics[height = 20mm, width = 70mm]{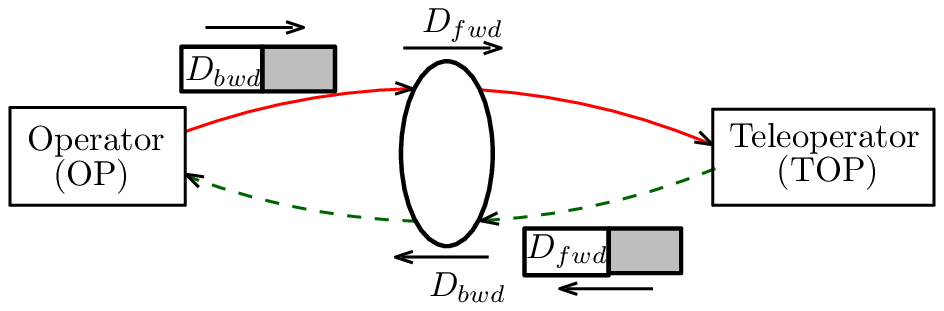}
    \caption{A schematic representation of the in-header delay notification mechanism. $D_{fwd}$ and $D_{bwd}$ indicate the end-to-end delays on the forward and backward channels, respectively.}
    \label{fig:delaynotif}
\end{minipage}
\hspace{0.55cm}
\begin{minipage}[b]{0.45\linewidth}
\centering
\includegraphics[height = 20mm, width = 50mm]{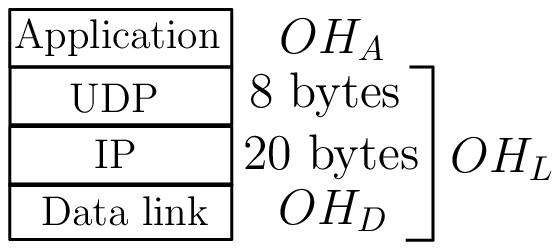}
\caption{Representation of the network protocol stack model, along
with the corresponding header size at each layer. The packet header structure of DPM is as described in Appendix~\ref{sec:applicationheader}.}
\label{fig:stack}
\end{minipage}
\end{figure}


We exploit the bidirectional nature of the telehaptic traffic to
convey end-to-end delays on each channel to the respective transmitter without
transmitting specialized reports (unlike RTP). The in-header delay
notification mechanism inserts the end-to-end delay encountered by the
latest received packet into the header of the packet to be
transmitted, as shown in Figure \ref{fig:delaynotif}. In particular,
the headers of packets transmitted on the forward channel include the
end-to-end delay experienced by the last received packet on the
backward channel, and vice-versa. This mechanism enables
real-time monitoring of the state of congestion on each channel separately, with
a negligible overhead of 3 bytes per packet.

The telehaptic nodes are time-synchronized using Network Time Protocol
(NTP) \cite{ref:ntp}. The end-to-end delay encountered by a telehaptic packet
received is thus calculated as the difference between the time of
reception and the timestamp of the earliest haptic sample embedded in
the received packet. Note that merging of multiple telehaptic fragments in a packet is explained in detail in Section \ref{subsubsec:networkbased}.

The in-header delay notification mechanism is more effective than the report-based notification
of RTP for three major reasons.  Firstly, the higher rate of delay
notifications provides finer details of network changes.
This enables the telehaptic nodes to swiftly adapt the telehaptic rate
to the changing network conditions.
Secondly, our scheme does not transmit specialized packets to convey
delay feedback, and thus induces a smaller overhead compared to
RTP. Thirdly, the in-header delay notification mechanism estimates the
delays on the forward and backward channels separately, enabling each
transmitter to adapt its rate based on the state of the corresponding
channel. 

\ignore{Thirdly, every node can carry out the end-to-end delay
  analysis associated with the channel on which it transmits data
  since the delay notification mechanism makes the end-to-end delay of
  the corresponding channel available to the node. On the other hand,
  analysis of the round trip delay possibly induces data rate
  adaptation at the operator due to variations in the backward channel
  and vice-versa, which is unnecessary. Therefore, the proposed
  feedback mechanism is effective even for asymmetric networks, which
  are more prevalent in real world.}

\subsubsection{Generation of rate-adaptation triggers}

Based on the trend observed in the measured delays on each channel, the network
feedback module generates two triggers for the corresponding
transmitter. The trigger $I_C$ signals that the channel is getting
congested; this causes the DPM module to reduce the telehaptic data
rate if possible (see Section~\ref{subsubsec:networkbased}). The trigger $I_S$ signals
that the channel delays are steady; this causes the DPM module to
probe if the channel has spare capacity by increasing the
telehaptic data rate if possible (see Section~\ref{subsubsec:networkbased}).



In order to trace the delay pattern, we use an exponentially weighted moving average
filter defined by
\begin{equation}
 d_{avg}(n) = \alpha*d(n) + (1-\alpha)*d_{avg}(n-1),
 \label{equ:mwafilter}
\end{equation}
where $0 < \alpha < 1.$ Here, $d(n)$ denotes the $n$\textsuperscript{th} end-to-end
delay measurement.\footnote{Note that the OP (TOP) may receive the
  same delay measurement multiple times; this can happen if the TOP
  (OP) makes multiple packet transmissions between successive
  receptions. To avoid the same delay measurement from resulting in
  multiple updates in Equation \eqref{equ:mwafilter}, we include a one-bit
  field named \textit{delay indicator} (field \textit{D} of Table
  \ref{table:packetstructure} in Appendix \ref{sec:applicationheader}) to the packet header. This field is set
  to 1 in case of a repetitive transmission of a previously computed
  delay, and 0 in case of the transmission of a newly computed delay.}

The network feedback module generates triggers as follows. The trigger
$I_C$ is generated on observing $N$ continuous increasing measurements
in $d_{avg}(\cdot).$ Note that a steady increase in the end-to-end
delays indicates that queues in the network are building up due to congestion. The
trigger $I_S$ is generated if the most recent $N$ entries in
$d_{avg}(\cdot)$ satisfy two conditions: (i) the entries exhibit
neither an increasing nor a decreasing trend, and (ii) the latter
$N-1$ entries are within a tolerance interval of 10\% around the
first. Note that generation of the trigger $I_S$ signals that network
conditions are steady. It is worth mentioning that since the generation of triggers is based on
a \emph{trend} of the end-to-end delays,
the proposed rate adaptation scheme remains robust to time synchronization errors of NTP.
In our experiments, reported in
Section~\ref{sec:results}, we set $\alpha =$ 0.2, as recommended in \cite{ref:montgomeryEWMA},
and $N =$ 8.

\subsection{Dynamic Packetization Module}
\label{subsubsec:networkbased}

In this section, we describe the dynamic packetization module (DPM), which
adapts the telehaptic data rate based on the triggers generated by the
network feedback module. We begin by presenting some calculations that
illustrate the extent of telehaptic data rate variation possible by
varying the packetization rate.

Assuming Ethernet on the data link layer, the overall overhead per
packet due to the link layer ($OH_D =$ 26 bytes), IP, and UDP headers
equals 54 bytes (see Figure~\ref{fig:stack}). Adding to this our
protocol's overhead ($OH_A$) of 13 bytes (see Appendix
\ref{sec:applicationheader}), we arrive at a net overhead of 67
bytes/packet. If we transmit each telehaptic fragment as a separate
packet (this corresponds to a 1 kHz packetization rate), this amounts
to an overall overhead rate of $OHR =$ 536 kbps. For a standard TOP
payload rate of 560 kbps (haptic - 96 kbps, audio - 64 kbps, video -
400 kbps), the overhead constitutes a substantial proportion (48.9\%)
of the telehaptic traffic.\footnote{The overhead represents an even
  higher proportion (72.09\%) of the telehaptic traffic from the OP to
  the TOP, since the payload is composed of only haptic data.}

\begin{figure}[!h]
\begin{minipage}[b]{0.45\linewidth}
\centering
\includegraphics[height = 40mm, width = 70mm]{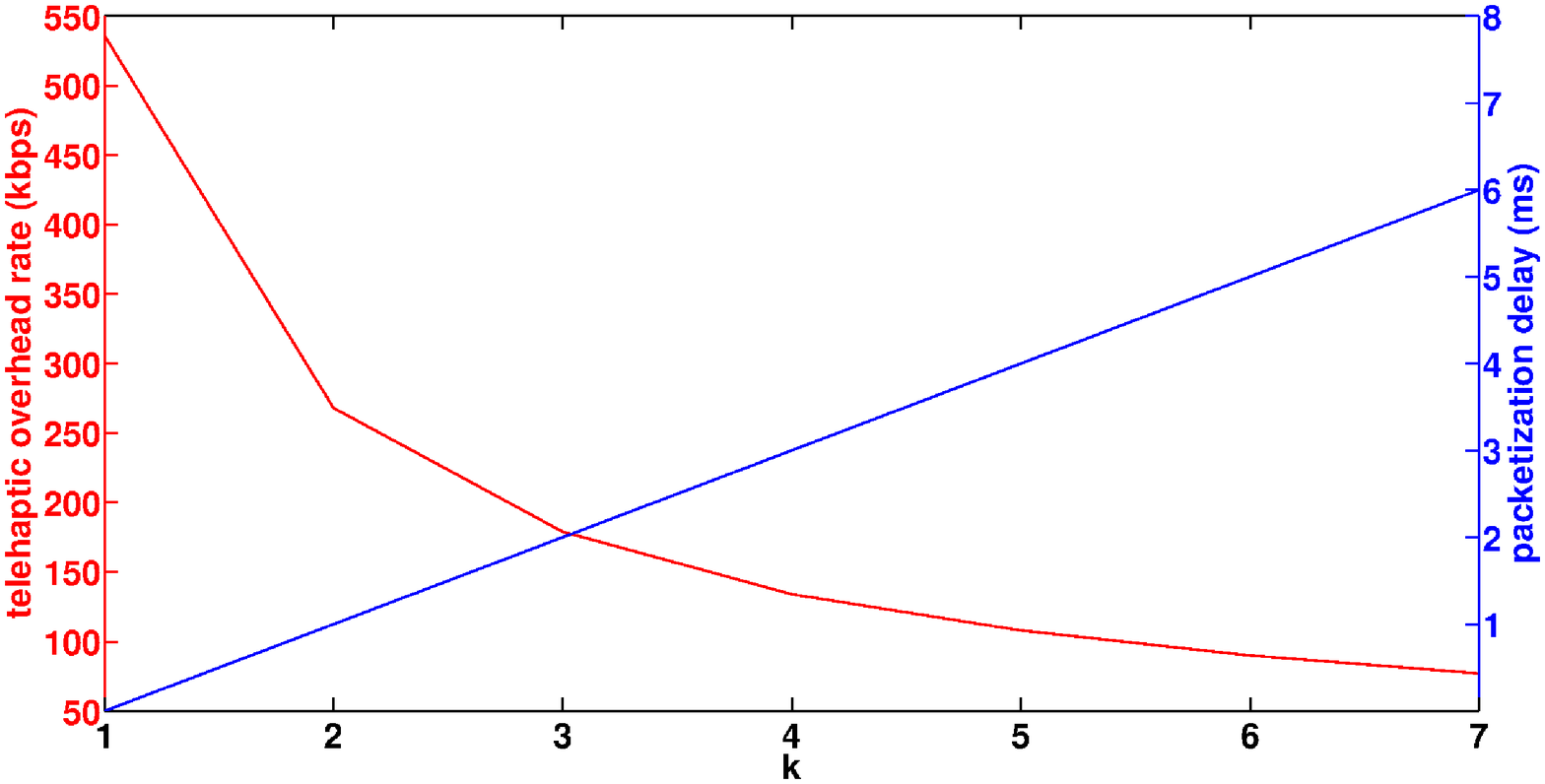}
\caption{Telehaptic overhead rate variation for different $k$-merge packetization schemes, along with the corresponding packetization delay.}
\label{fig:kplot}
\end{minipage}
\hspace{0.55cm}
\begin{minipage}[b]{0.45\linewidth}
\centering
\includegraphics[height = 40mm, width = 60mm]{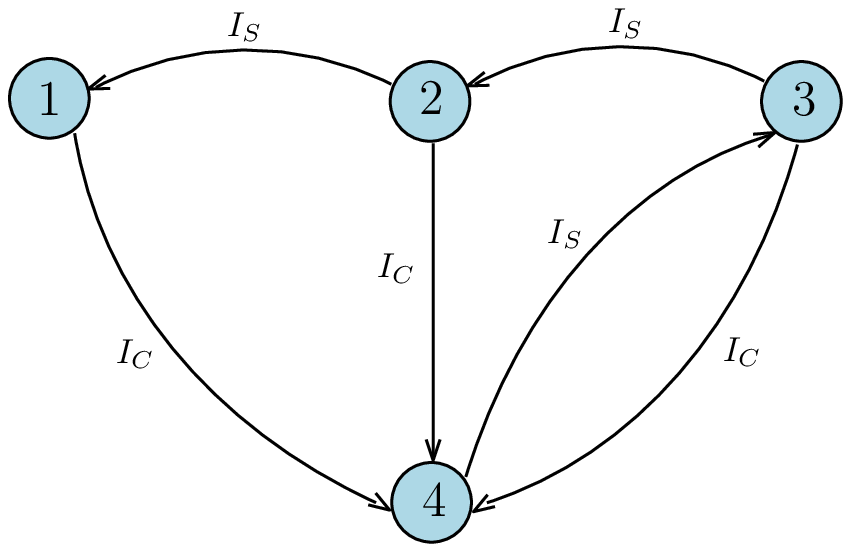}
\caption{A finite state transition diagram representation of the step-increase-multistep-decrease approach of DPM with $k_{max}$ = 4.}
\label{fig:statediagram}
\end{minipage}
\end{figure}

Now, suppose that we merge $k$ consecutive telehaptic fragments into a
single packet for transmission. We refer to this scheme as the
\emph{$k$-merge} packetization scheme, and we refer to the special
case $k = 1$ as the \emph{no-merge} packetization scheme. The
telehaptic data rate $R_k$ corresponding to the $k$-merge
packetization scheme (in kbps) is given by
\begin{equation}
 R_k = D + \frac{OHR}{k}
 \label{equ:rk}
\end{equation}
where $D$ is the telehaptic payload generation rate
given by \eqref{equ:payloadrate} and $OHR$ denotes the overhead rate
under the no-merge scheme. Taking $OHR = $ 536 kbps, Figure
\ref{fig:kplot} presents the variation of telehaptic overhead rates
and packetization delay for different $k$-merge schemes. Note that these packetization delays
correspond to the earliest haptic sample in the packet. Assuming $D =$
560 kbps, we see that on the backward channel the telehaptic transmission rate for the
no-merge scheme equals 1096 kbps, whereas the transmission
rate for the 4-merge scheme equals 694 kbps. We observe that there is a
substantial scope for losslessly varying the telehaptic transmission
rate by controlling the packetization parameter $k.$ Of course, the
data rate reduction from increasing $k$ comes at the cost of a higher
packetization delay at the source.

The idea behind DPM is to dynamically adapt the packetization
parameter $k$ depending on network conditions. In other words, DPM
dynamically switches between different $k$-merge schemes based on the
triggers from the network feedback module. From
Figure~\ref{fig:kplot}, we note that the overhead reduction becomes
insignificant for large values of $k,$ whereas the packetization delay
grows linearly in $k.$ Thus, DPM confines the adaptation of $k$ to the
range $1 \leq k \leq k_{max}.$ In this work, we set $k_{max} = 4.$

DPM is a \textit{step-increase-multistep-decrease} (SIMD)
algorithm. This is a variation of the classical
additive-increase-multiplicative-decrease (AIMD) congestion control
mechanism of TCP \cite{ref:chiuaimd}. Specifically, on receiving the
trigger $I_C$ (recall that this trigger signals that the network is
getting congested), DPM sets $k = k_{max}.$ Thus, on sensing
congestion in the network, DPM decreases the telehaptic data rate
aggressively in order to decongest the network in the shortest
possible time. On the other hand, on receiving the trigger $I_S$
(recall that this trigger signals that network delays are steady), DPM
decreases $k$ by 1 if $k >$ 1. Thus, on sensing that the network
is in a steady state, DPM probes if a higher data rate is achievable
by decreasing $k$ by one unit. Figure \ref{fig:statediagram} shows a
finite state transition diagram representation of
DPM. 

\begin{figure}[!h]
\begin{center}
\includegraphics[height = 75mm, width = 110mm]{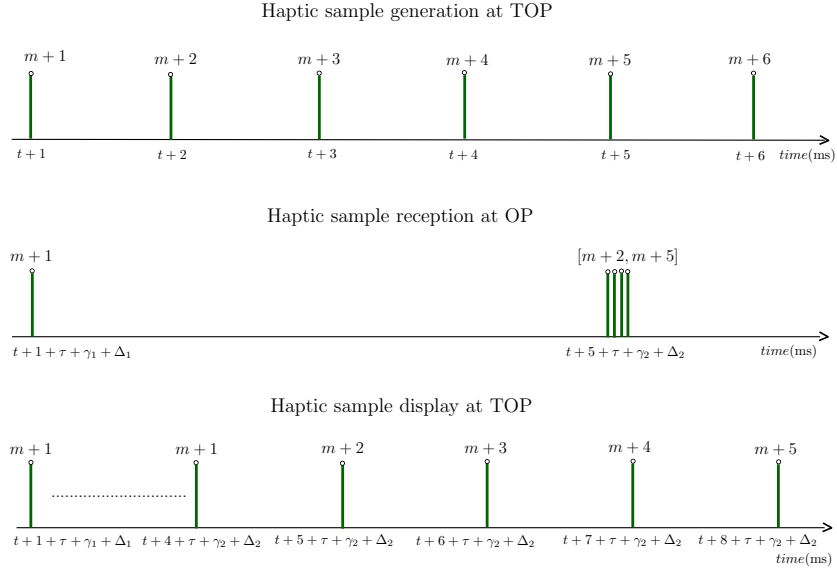}
\caption{Timing diagram illustrating haptic sample transmission at TOP, reception and display at OP using zero-order hold strategy. The samples bunched together indicate simultaneous reception due to the 4-merge packet.}
\label{fig:kmergeexample}
\end{center}
\end{figure}

Note that DPM's dynamic packet rate adaptation will induce additional
jitter in the receiver. To get sense of the jitter caused by DPM, we
perform the following simple analysis, focusing only on haptic jitter
(note that the haptic stream has the tightest jitter constraint). It
is easy to see that the maximum jitter occurs when switching from
$k=1$ to $k = k_{max} = 4.$ Consider the sequence of haptic samples
shown in Figure~\ref{fig:kmergeexample}. Suppose that initially,
$k=1.$ Note that sample number $m+1$ is generated at time $t+1,$ and
is received and displayed at time $t+1+\tau+\gamma_1+\Delta_1$. Here, $\tau$
denotes the one-way propagation delay, and $\Delta_1$ and $\gamma_1$ denote the queueing delay and transmission delay
of the packet containing sample number $m+1$, respectively. Now, suppose that
starting from sample number~$m+2,$ we switch from $k=1$ to $k = 4.$ In
this case, sample $m+2,$ which is generated at time $t+2,$ will only
get transmitted at time $t+5$ (along with the next three samples), and
will get received and displayed at time $t+5 + \tau+\gamma_2+\Delta_2.$ Here,
$\Delta_2$ and $\gamma_2$ denote the queueing and transmission delays,
respectively, experienced by the packet containing sample $m+2.$ Thus,
the jitter of the haptic sample $m+2$ equals the difference between
its actual display time and its expected display time:
$(t+5 + \tau+\gamma_2+\Delta_2) - (t+2+\tau+\gamma_1+\Delta_1) = 3 + (\Delta_2-\Delta_1) + (\gamma_2- \gamma_1).$
Note that $\gamma_2 \leq 4 \gamma_1.$ Assuming then that $\Delta_1$ and
$\Delta_2$ are comparable, we can bound the jitter by $3(1+\gamma_1).$
Thus, we see that by restricting $k$ to be at most 4 under DPM, we
introduce an additional jitter of at most $3(1+\gamma_1)$ on the haptic
stream. Note that the subsequent 4-merge packet (carrying haptic samples [$m+6, m+9$]) arrives at $t+9 + \tau+\gamma_2+\Delta_2$. We validate the correctness
of the jitter analysis through simulation results reported in Table \ref{table:delaysummary}.\\

\begin{figure}[!h]
\begin{center}
\includegraphics[height = 65mm, width = 110mm]{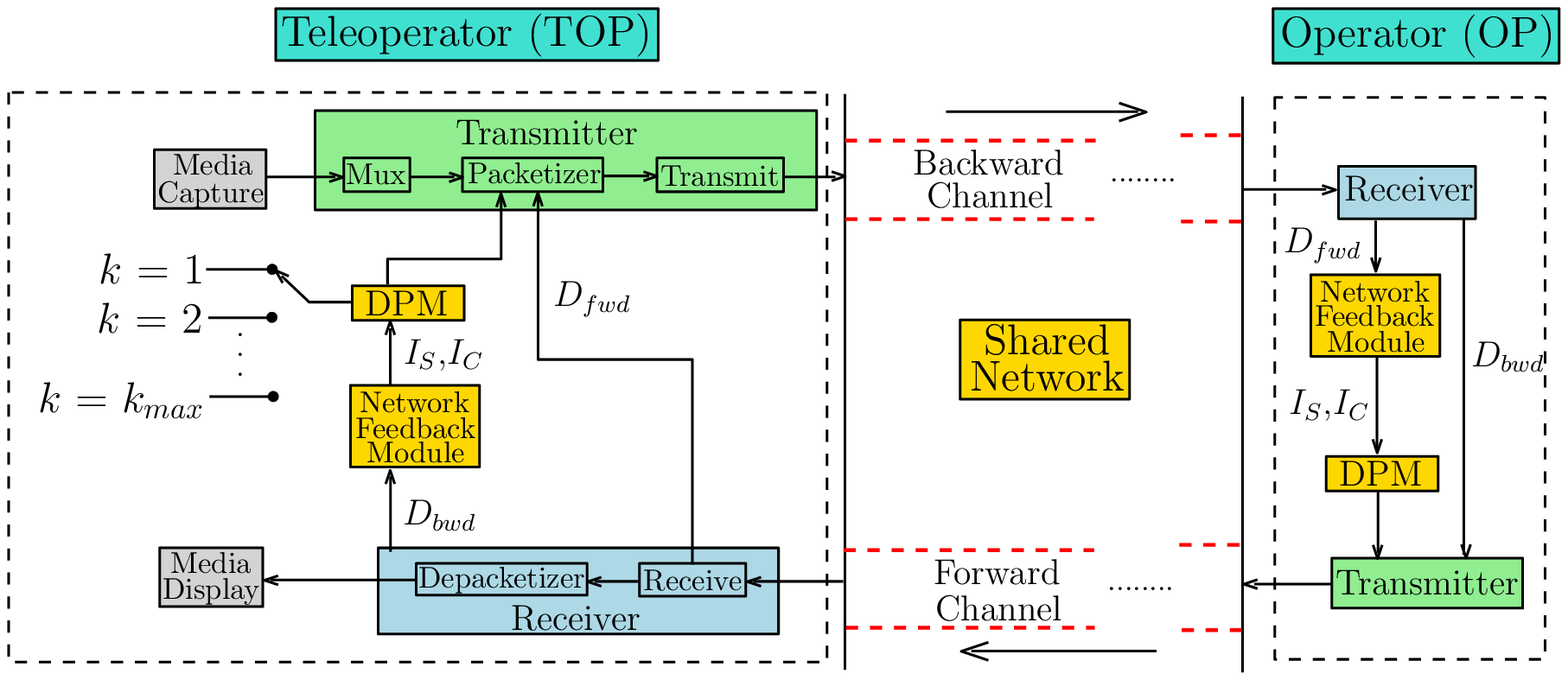}
\caption{A block diagram showing the architecture of the proposed telehaptic communication framework. The design at the OP is similar to that of the TOP, and is not shown for brevity.}
\label{fig:architecture}
\end{center}
\end{figure}

\noindent \textbf{Overview of Protocol Architecture:} Figure
\ref{fig:architecture} presents an overview of the proposed telehaptic
communication framework. We explain the working with respect to the
TOP, whereas similar operations are carried out at the OP as well. On
receiving the telehaptic packet at the TOP, the \textit{depacketizer}
module decodes the header information. Based on the header contents,
the payload is forwarded to the appropriate media display devices. The
backward channel delay ($D_{bwd}$) in the header is supplied to the
\textit{network feedback module} for learning the recent changes in
the backward channel. Based on the delay analysis, the network
feedback module generates triggers ($I_S, I_C$) appropriately. On
arrival of a trigger, the DPM selects $k$, which is communicated to
the \textit{packetizer} for composing the telehaptic packets. The TOP
also calculates the end-to-end delay on the forward channel
($D_{fwd}$) after every packet reception, which is sent to the
\textit{packetizer} for inclusion in the packet header that is transmitted to the OP. \\


%

\section{Experimental Design}
\label{sec:exptdesign}

In this section, we describe the setup used in our experiments to
assess the performance of the telehaptic data transmission scheme
proposed in this paper. The objective of the experiments is to investigate 
the ability of DPM to perform congestion control under heavy cross-traffic scenarios.
The performance metrics we consider are QoS adherence, signal-to-noise ratio (SNR)
of the reconstructed haptic signal at the receiver, and
the perceptual quality of the displayed haptic-audio-video signal.
We first describe the setup used in our
simulations, and then describe the setup of the real-time telepottery
experiment. The results of these experiments follow in
Section~\ref{sec:results}.

\subsection{Simulation Setup}
\label{subsec:simulationsetup}
\begin{figure}[!h]
\begin{center}
\includegraphics[height = 38mm, width = 80mm]{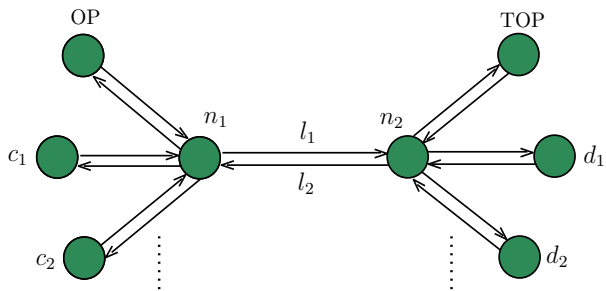}
\caption{Single bottleneck dumbbell network topology design for simulation of the telehaptic communication.}
\label{fig:topology}
\end{center}
\end{figure}

Our simulations are carried out using NS-3, a discrete event network
simulator \cite{ref:ns3}. We consider a network with a single
bottleneck dumbbell topology connecting the OP and the TOP, as shown
in Figure~\ref{fig:topology}. In order to simulate asymmetric
network conditions on the forward and the backward channels,
we create unidirectional links between the OP and the TOP node. All links have identical capacities (denoted by $\mu$) of
1.5 Mbps.\footnote{1.5 Mbps has been picked to represent the typical capacity of a medium speed internet
link. However, the nature of our findings remain
robust to the channel capacity.} To simulate cross-traffic on the forward (respectively,
backward) channel, we add source-destination pairs $(c_i,d_i)$
(respectively, $(d_j,c_j)$) as indicated in
Figure~\ref{fig:topology}. Note that $l_1$ and $l_2$ act as the
bottleneck links for the telehaptic traffic on the forward and
backward channels, respectively. Thus, queueing delay experienced by
the telehaptic application due to network cross-traffic is observable
only at the intermediate nodes $n_1$ and $n_2$. 

The haptic payload rates on the forward and backward channels are set
to 192 kbps and 96 kbps \cite{ref:vineetncc},
respectively. The TOP generates audio frames of size 160
  bytes 20 ms apart, and video frames of size 2 kB, 40 ms apart. This
  corresponds to payload rates on the backward channel of 64 kbps and
  400 kbps respectively for audio and video.
Considering the application layer header sizes (see Appendix
\ref{sec:applicationheader}), the no-merge data rate on the forward
and backward channels are calculated to be 688 kbps and 1096 kbps,
respectively.

Finally, the propagation delay of each link is set to 5 ms. Hence, the
one-way propagation delay (denoted by $\tau$) is 15 ms,
which is typically the propagation delay exhibited by a
transcontinental link of around 2000 miles. All nodes follow
first-in-first-out (FIFO) and droptail queueing of packets.

\ignore{
(TO EDIT.)$c_1$ and $d_1$ are the CBR cross-traffic sources, whereas
$c_2$ and $d_2$ represent other variable bitrate (VBR) traffic sources
in the network.

(TO EDIT.)For brevity, we mention the cross-traffic conditions on the
backward channel only. The VBR cross-traffic intensity, that is from
$d_2$ to $c_2$, $R_{vbr} \in$ [320, 480] kbps with a mean of 400
kbps. Under the aforementioned network settings, the target range of
CBR cross-traffic $R_{cbr}$, from $d_1$ to $c_1$, for which the
$k-$merge algorithm results in an effective congestion control [13,
408] kbps\footnote{$R_1 + R_{vbr} + R_{cbr} >$ 1300, and $R_4 +
  R_{vbr} + R_{cbr} \leq$ 1300. It is to be noted that in evaluating
  the above conditions, we consider the mean value of the VBR
  traffic}. The various simulations presented use different $R_{cbr}$,
as described in Section \ref{sec:results}. The VBR cross-traffic is
active throughout the length of the experiments. $R_{cbr}$ is switched
on at 500 ms, and stays active throughout thereafter, unless mentioned
otherwise.
}
\subsection{Perceptual Experiment Setup}
\label{subsec:perceptsetup}

\begin{figure}[!t]
\centering
  \subfloat[]{\includegraphics[height = 45mm, width = 60mm]{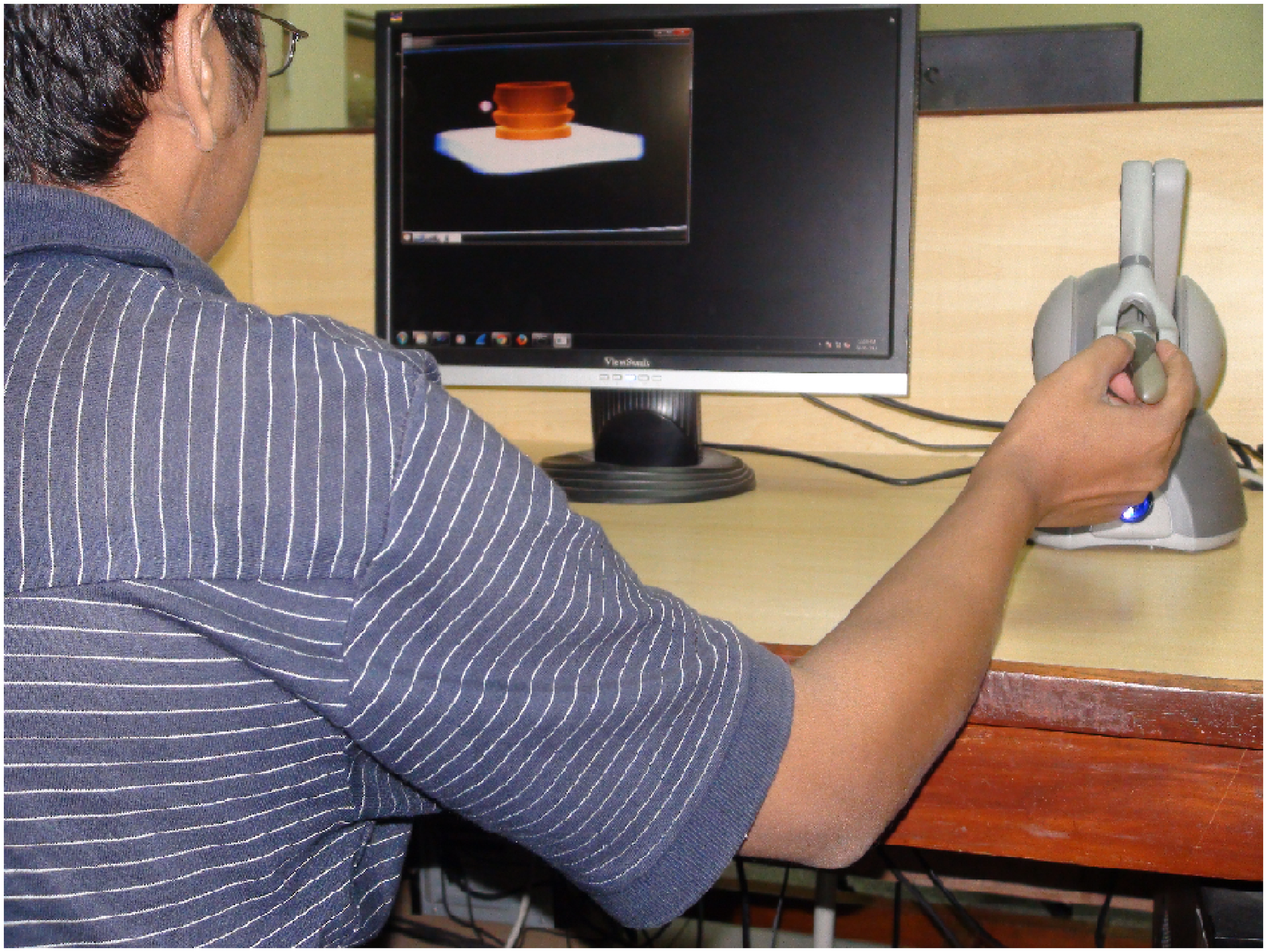} \label{fig:operator}}~
  \subfloat[]{\includegraphics[height = 45mm, width = 60mm]{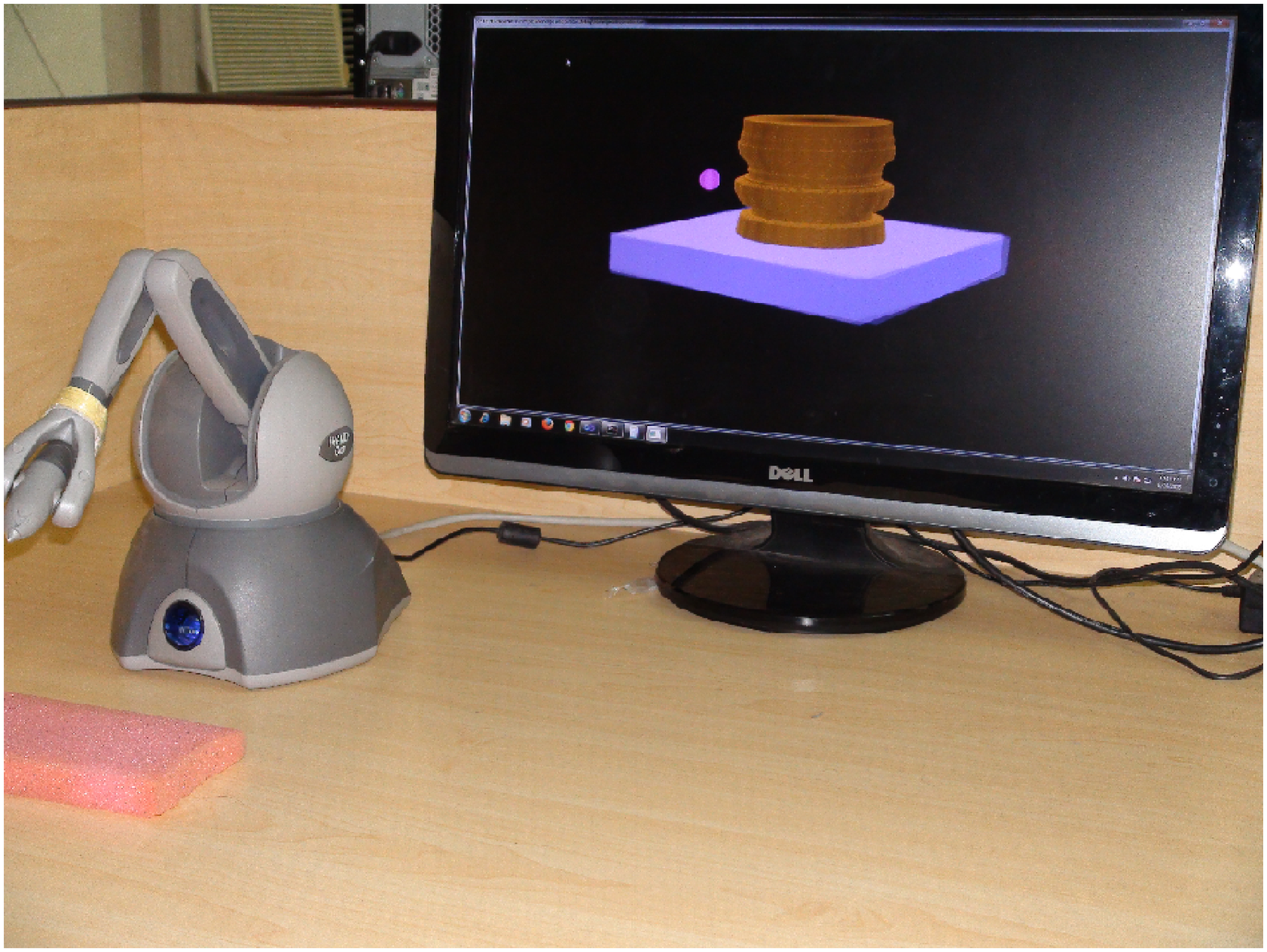}\label{fig:teleoperator}}
\caption{Real-time telepottery experimental setup showing (a) human operator (b) teleoperator.}
\label{fig:telepottery}
\end{figure}


It is important to investigate the qualitative effect of DPM on human
multimedia perception, which is not possible through simulations. For
this purpose, we conduct a real-time telepottery experiment in which a
human subject interacts with a remote, virtual pottery model on a real
network through haptic, audio and visual feedback, as described in
\cite{ref:vineetncc}. Figure~\ref{fig:telepottery} demonstrates the
setup of the telepottery experiment showing the human subject remotely
exploring the virtual clay model. The volume preserving pottery model
\cite{ref:potterychaudhury} is rendered at the TOP, which is equipped
with a haptic device and a generic webcam. The interaction with the
remote scene happens through audio-visual feedback and a separate
haptic device for the haptic feedback. The master-slave relationship
between the two haptic devices is implemented using a
proportional-derivative controller; see \cite{ref:vineetncc} for more
details.

The subjects were initially briefed about the concept of force
feedback as few of them were new to the notion of haptics. Later, we
explained them the telepottery task in detail accompanied by a live
demonstration of the task. The telepottery task involves
  the subject exploring and manipulating a rotating virtual clay
  model.  The task is to design a clay pot. There is no benchmarking
  so far as the shape of the pots is concerned, since the idea behind
  the experiment is to assess human perception and not skill. The
  subject pushes the haptic device stylus so as to establish contact
  with the clay model and shape it into a pot. The \textit{training}
phase involved the participants performing the task to get acquainted
with the telepottery setup. During this phase, the participants
explored the telepottery model under an expert's guidance until they
were confident of performing the task independently.  In order to
avoid any perceptual degradation due to the network, the training was
performed on a very high bandwidth network, under the no-merge
packetization scheme.

After the training, the subjects were moved to a \textit{test} setup
consisting of a network emulator tool that allows for configuring the
network capacity and propagation delay. Under the emulated network
conditions, the subjects independently perform the telepottery task
twice: once with no-merge scheme, and once with the proposed DPM
scheme. Finally, the subjects were asked to grade the experience of
each of the two test experiments, relative to the training, based on
three perceptual parameters: \textit{transparency} (the subjects felt
as if they were present in the remote virtual environment and are
directly interacting with the objects)
\cite{ref:lawrencetransparency}, \textit{smoothness} (how smooth or
jerky is the feedback) \cite{ref:isomurasmoothness} and
\textit{overall experience}.  The grading of each of the three
parameters was based on degradation category rating (DCR)
\cite{ref:hoshinoolfactory,ref:suzukibilateral,ref:fujimotoerror} that
assigns
a subjective scale to a text descriptor in the following manner:\\
5- imperceptible; 4- slight disturbance, but not annoying;
3- slightly annoying; 2- annoying; 1- very annoying.\\
For example, the subjects chose 5 if they felt that the degradation in
perceptual quality of the test experiments was imperceptible compared
to the training phase.

The average training duration was measured to be around 12 minutes,
and the average duration for each of the test experiments was around 6
minutes. The subjects had no prior knowledge about the protocol being
tested, thereby avoiding grading bias.


\subsubsection{System Settings}
\label{subsubsec:netwsetting}
In the real-time telepottery experiment, we use two Phantom Omni
haptic devices which capture the low-level dimensions of human haptic
perceptual system \cite{ref:carbonDimensions}. Two desktop computers,
each with 4 GB RAM and running Windows 7 operating system are
employed. The audio-visual information is captured at the TOP using a
Microsoft Lifecam VX-2000 webcam. The TOP transmits
  uncompressed audio and video frames at the rate of 164 kbps and 1.1
  Mbps, respectively.
  For these experiments, we increase the channel capacity by 800 kbps
  compared to the simulation setup to account for the additional
  audio/video payload.

  The network is emulated using a standard network
    emulator tool called Dummynet \cite{ref:dummynet}. The training
  phase of the telepottery experiment is performed on a 100 Mbps
  network. For the testing phase, the emulated channel capacity and
  one-way propagation delay are configured to 2.3 Mbps and 15 ms,
  respectively, for both the forward and the backward channel. In the
  testing phase, we introduce constant bit-rate (CBR) cross-traffic
  stream of intensity 400 kbps on the backward channel. In addition,
  we introduce variable bit-rate (VBR) source with intensity $R_{vbr}
  \in$ [320, 480] kbps with a mean of 400 kbps on the backward
  channel.

\subsubsection{Human Subjects}
\label{subsubsec:humansubjects}
The call for participation in the telepottery task was published on noticeboards in
the university. All human subjects who took part in the experiment were
either students or faculty members at the university.
A total of twenty subjects (ten female and ten male, eighteen right-handed
and two left-handed) participated in the perceptual task. The subjects belonged to the age
group of twenty three to fifty two years, and none of them suffered from any known
neurophysiological disorders. Out of the twenty participants, 
fourteen were novice haptic users and the rest were regular users of haptic devices.
However, all subjects underwent extensive training prior
to the test experiments.

\section{Experimental Results}
\label{sec:results}

In this section, we present a comprehensive experimental evaluation of
DPM. Simulation results are presented in
Section~\ref{subsec:simresults}, and the results of our perceptual
experiments are presented in
Section~\ref{subsec:percepexptresults}. In
Section~\ref{sec:comparison}, we compare the performance of DPM with
the state of the art in telehaptic communication protocols.

\subsection{Simulation Results}
\label{subsec:simresults}

In this section, we present the performance evaluation of DPM via
simulations. Specifically, we analyze the interplay between DPM and
network-oblivious cross-traffic, highlighting DPM's response to highly
congested network conditions. For brevity, we present results
corresponding to only the backward channel; the performance of DPM on
the forward channel is similar.


The simulation begins at time $t = 0,$ at which point the telehaptic
stream commences transmission. Starting at $t = 0,$ we also maintain
VBR stream on backward channel with intensity $R_{vbr} \in$ [320, 480]
kbps with a mean of 400 kbps.\footnote{A Skype video-conferencing
  connection consumes approximately 400 kbps of bandwidth in each
  direction. Thus, the VBR cross-traffic can be thought of as a
  video-conferencing stream contending with the telehaptic stream on
  the bottleneck link.} At $t=500$ ms, we additionally introduce CBR
cross-traffic stream on the backward channel.  The intensity of the
CBR source is used as a control parameter to tune the level of
congestion on the backward channel. Note that the peak telehaptic data
rate on the backward channel equals 1096 kbps (under no-merge
packetization), whereas the minimum data rate equals 694 kbps (with
$k=4$ under DPM). Thus, denoting the intensity of the CBR stream on
the backward channel by $R_{cbr},$ we note that when $R_{cbr} > 4$
kbps, the telehaptic stream has insufficient bandwidth to transmit at
its peak rate. (Recall that $\mu =$ 1.5 Mbps.) Moreover, $R_{cbr} >
406$ kbps implies that the network is overloaded, since the available
capacity is insufficient to even sustain the minimum telehaptic data
rate. Thus, the effectiveness of DPM is to be gauged over the range of
$R_{cbr} \in [4,406]$ kbps. In most of our experiments, we set
$R_{cbr} = 400$ kbps, which represents a highly congested backward
channel. The cross-traffic rates on the forward channel are
identical to that on the backward channel.
The simulations run for 500 seconds. The throughput, average
jitter and packet loss measurements presented in this section are
computed after the CBR cross-traffic is switched on, i.e., over the
interval $t \in$ [0.5, 500] seconds.  

It is important to note that since the proposed protocol
  operates at the transport layer (TL), all delays reported in this
  section are TL-TL measurements. In other words, we report the
  latency between the arrival of a haptic/audio/video sample at the TL
  of the sender and the reception of the same sample at the TL of the
  receiver. The delays experienced by the OP/TOP in practice (the
  so-called `glass-to-glass' delays) would include the additional lag
  introduced by the media devices as well as the encoding/decoding
  latency.\\

\begin{figure}[!h]
\centering
  \subfloat[]{\includegraphics[height = 40mm, width = 73mm]{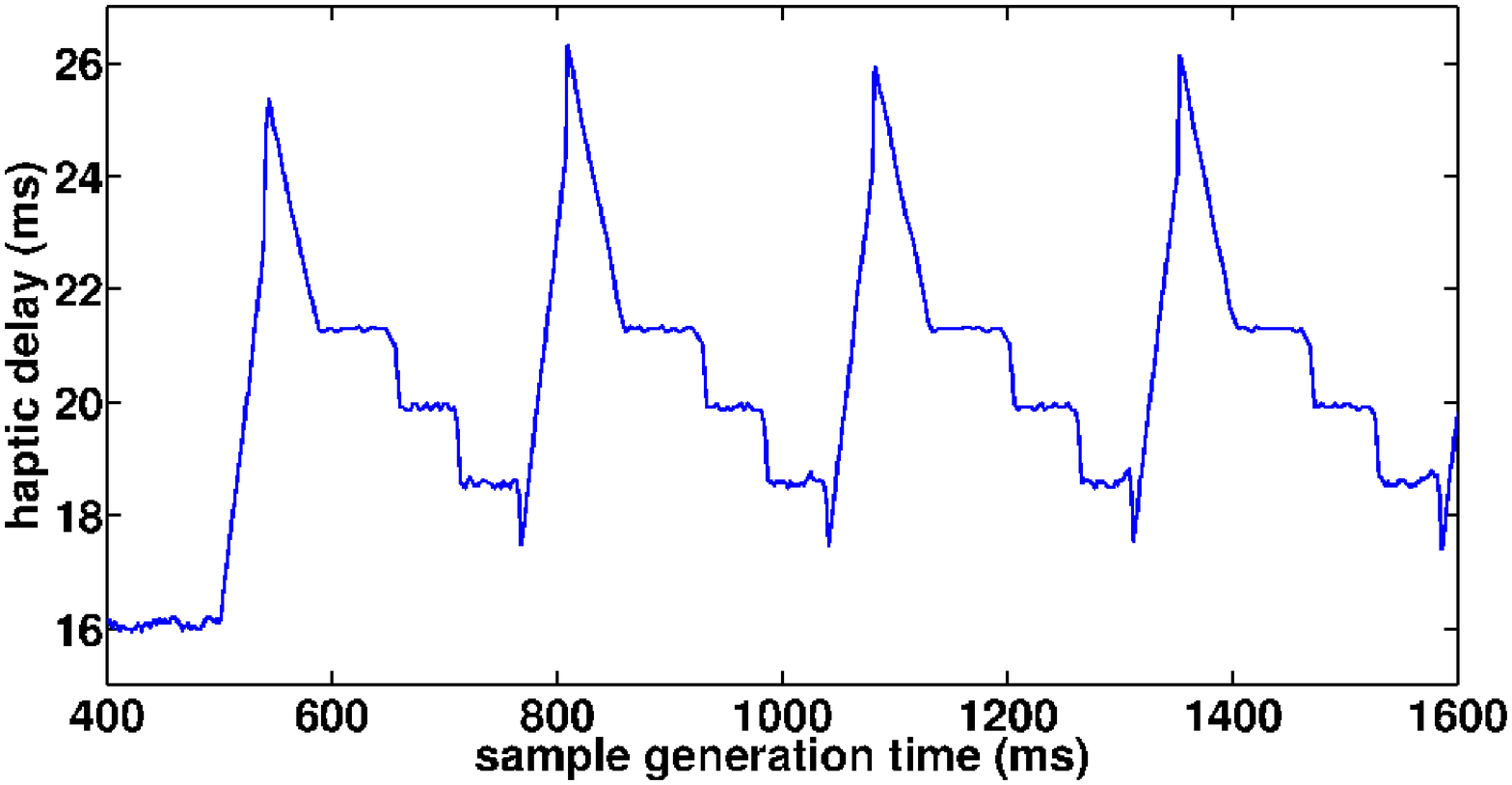} \label{fig:hapdelay2}}~ \hspace{-0.8cm}  
  \subfloat[]{\includegraphics[height = 40mm, width = 73mm]{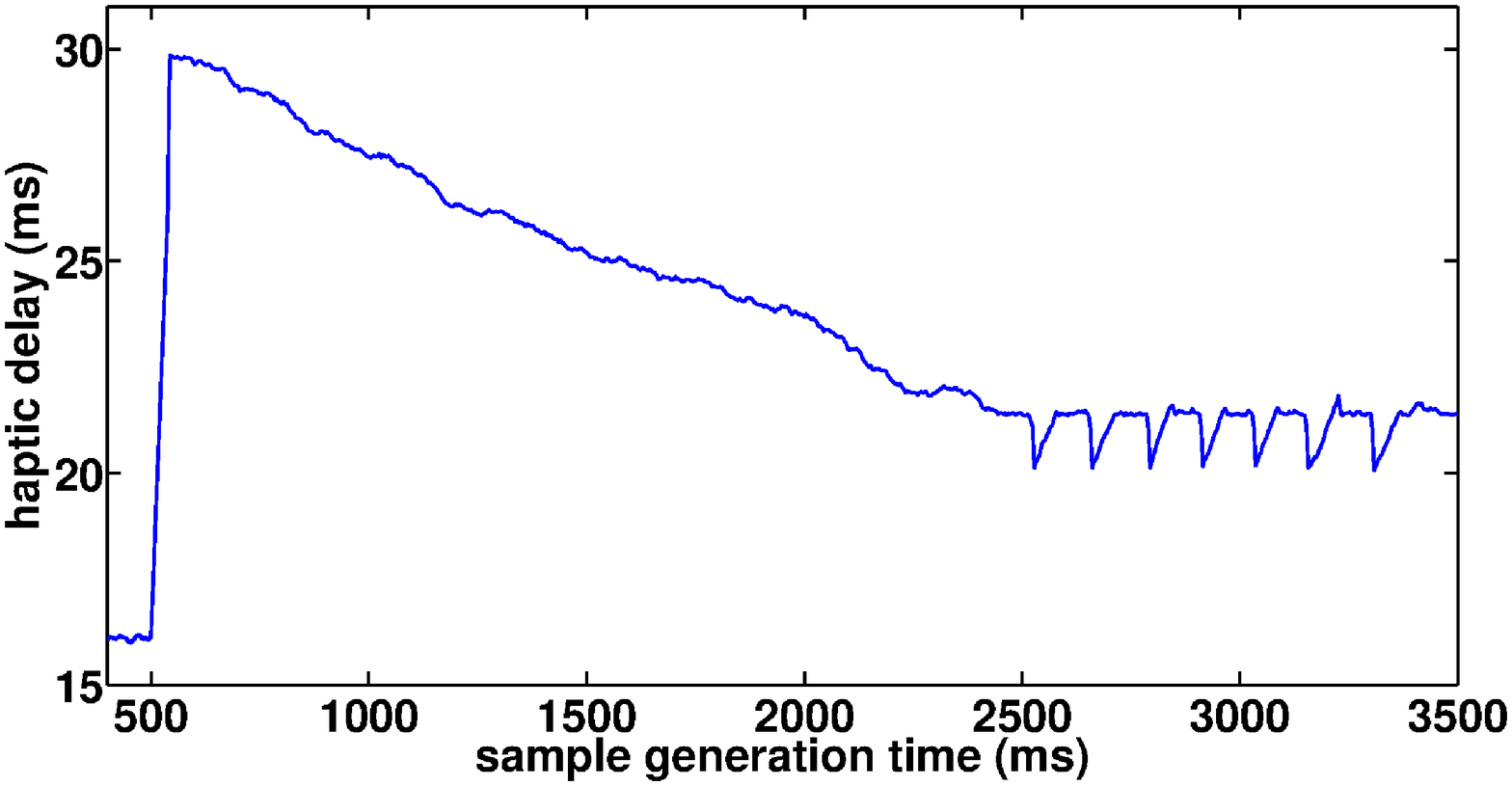} \label{fig:hapdelay4}}
\caption{Haptic delay profile as a result of DPM in presence of CBR cross-traffic (a) 260 kbps (b) 400 kbps.}
\label{fig:delayzones}
\end{figure}

\noindent \textbf{Temporal variation of telehaptic delay:} We begin by
demonstrating the temporal evolution of the delay experienced by the
telehaptic stream under DPM. Figure~\ref{fig:delayzones} shows the
delay experienced by the haptic samples as a function of the sample
generation time, corresponding to $R_{cbr}$ = 260 kbps and 400 kbps.

Let us first consider Figure~\ref{fig:delayzones}(a). For $R_{cbr}$ =
260 kbps, the capacity available to the telehaptic stream on the
backward channel equals 840 kbps, which is less than the no-merge
transmission rate of 1096 kbps, but more than the 2-merge transmission
rate of 828 kbps. Once the CBR source turns on at $t =$ 500 ms, the
telehaptic stream, initially operating at $k = 1,$ sees a rapid delay
build-up. DPM responds to this build-up by switching to $k = 4.$ This
aggressive rate reduction allows the network buffers to drain quickly,
avoiding a QoS violation. Once DPM sees a steady delay zone, it probes
for a higher telehaptic data rate by decreasing $k$ by 1. But when DPM
makes the switch from $k = 2$ to $k = 1,$ the overall network load once again exceeds the capacity of the bottleneck
link. This in turn leads to a delay build-up, and the cycle repeats.

Figure~\ref{fig:delayzones}(b) has a similar interpretation. For
$R_{cbr}$ = 400 kbps, the capacity available to the telehaptic stream
on the bottleneck link equals 700 kbps, which is less than the 3-merge
transmission rate of 739 kbps, but more than the 4-merge transmission
rate of 694 kbps. In this case, the switch from $k = 4$ to $k = 3$
causes a delay build-up, forcing DPM to revert to $k = 4.$

In conclusion, we see that DPM adapts its transmission rate depending
on the intensity of cross-traffic it experiences. Moreover, against a
steady cross-traffic, DPM results a roughly periodic delay
evolution. This is typical of congestion control algorithms; see, for
example, \cite{ref:cubic}. Note that even when the backward channel is
highly congested (see Figure~\ref{fig:delayzones}(b)), DPM manages to
keep the telehaptic delays below the prescribed QoS limits.\footnote{
  It is worth remarking that the haptic delay is dependent
    on the overall cross traffic intensity, the link capacities, as
    well as propagation delays. For haptic QoS compliance, we need to
    ensure that the maximum haptic delay does not exceed 30 ms. We
    perform a mathematical analysis for characterizing the maximum
    haptic delay in Appendix~\ref{sec:upperbound}. This enables us to
    identify the network settings under which haptic QoS adherence is
    feasible. Furthermore, we extend this characterization to audio
    and video in Appendix~\ref{sec:AVbound}, and show that the haptic
    QoS compliance in general guarantees audio and video QoS
    compliance.}  \vspace{-0.5cm}
\begin{figure}[!h]
\begin{center}
\includegraphics[height = 45mm, width = 120mm]{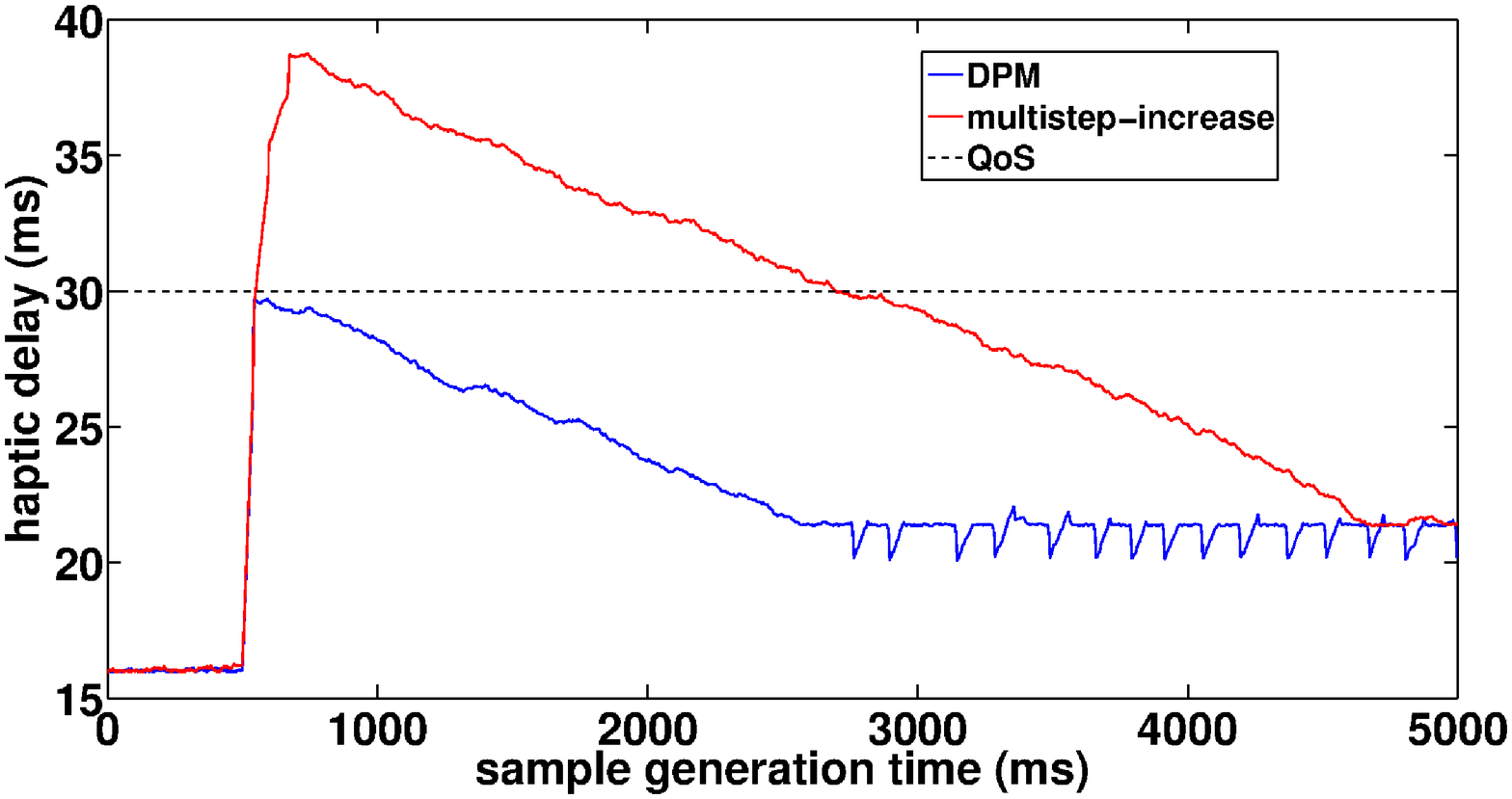}
\caption{Haptic delay variation with DPM and multistep-increase approaches, in presence of $R_{cbr}$ = 400 kbps.}
\label{fig:stepVSmultistep}
\end{center}
\end{figure}

\vspace{-0.5cm}
\noindent \textbf{Benefits of step-increase in DPM:} Recall that DPM
responds to network congestion with an aggressive transmission rate
reduction (achieved by a \emph{step-increase} in $k$ to $k_{max}$), as
opposed to a gradual transmission rate reduction (which would be
achieved by a \emph{multistep-increase} in
$k$). Figure~\ref{fig:stepVSmultistep} highlights the benefits of
employing the step-increase mechanism over a multistep-increase
approach for telehaptic data rate reduction. Specifically, we compare
the performance of DPM with an algorithm that increases $k$ by
one on receiving the congestion trigger $I_C$ (so long as $k <
k_{max}$). For this experiment, we set $R_{cbr}$ = 400 kbps. Once
the CBR stream starts, the telehaptic stream, initially operating at
$k =1,$ experiences a rapid delay build-up due to increased queueing
in the network. Note that DPM responds with an aggressive rate
reduction ($k = 4$), allowing the network buffers to get flushed
quickly, avoiding a QoS violation. On the other hand, the
multistep-increase approach cuts the transmission rate in stages,
requiring three rate adaptations before setting $k = 4$. As a result,
network decongestion occurs much later, leading to a violation of the
haptic QoS constraint. Thus, we conclude that DPM's SIMD approach is
suitable for congestion control for delay-critical telehaptic
applications.\\

\ignore{
  We conducted the simulations with several cross traffic
  intensities. For brevity, we report the results for $R_{cbr}$ = 380
  kbps, i.e. $k_{opt}$ = 4. The measurements demonstrate that the
  step-increase results in a faster congestion control, in addition to
  satisfying the QoS needs, whereas the multistep-increase readjusts
  the telehaptic data rate thrice causing the haptic delay to exceed
  the QoS threshold from $t_f$ = 536 ms to 1000 ms. The effect of
  change in the telehaptic data rate (in case of multistep approach)
  to $R_2$, $R_3$ and $R_4$ is observed as shift in the delay slope at
  536 ms, 600 ms and 684 ms, respectively. The graphs also suggest
  that the single-step approach reduces the haptic delay vulnerability
  to QoS violations, in case of further cross traffic additions, since
  it begins network decongestion at an earlier time instant than the
  multistep method.}

\begin{figure}[t]
\begin{center}
\includegraphics[height = 45mm, width = 120mm]{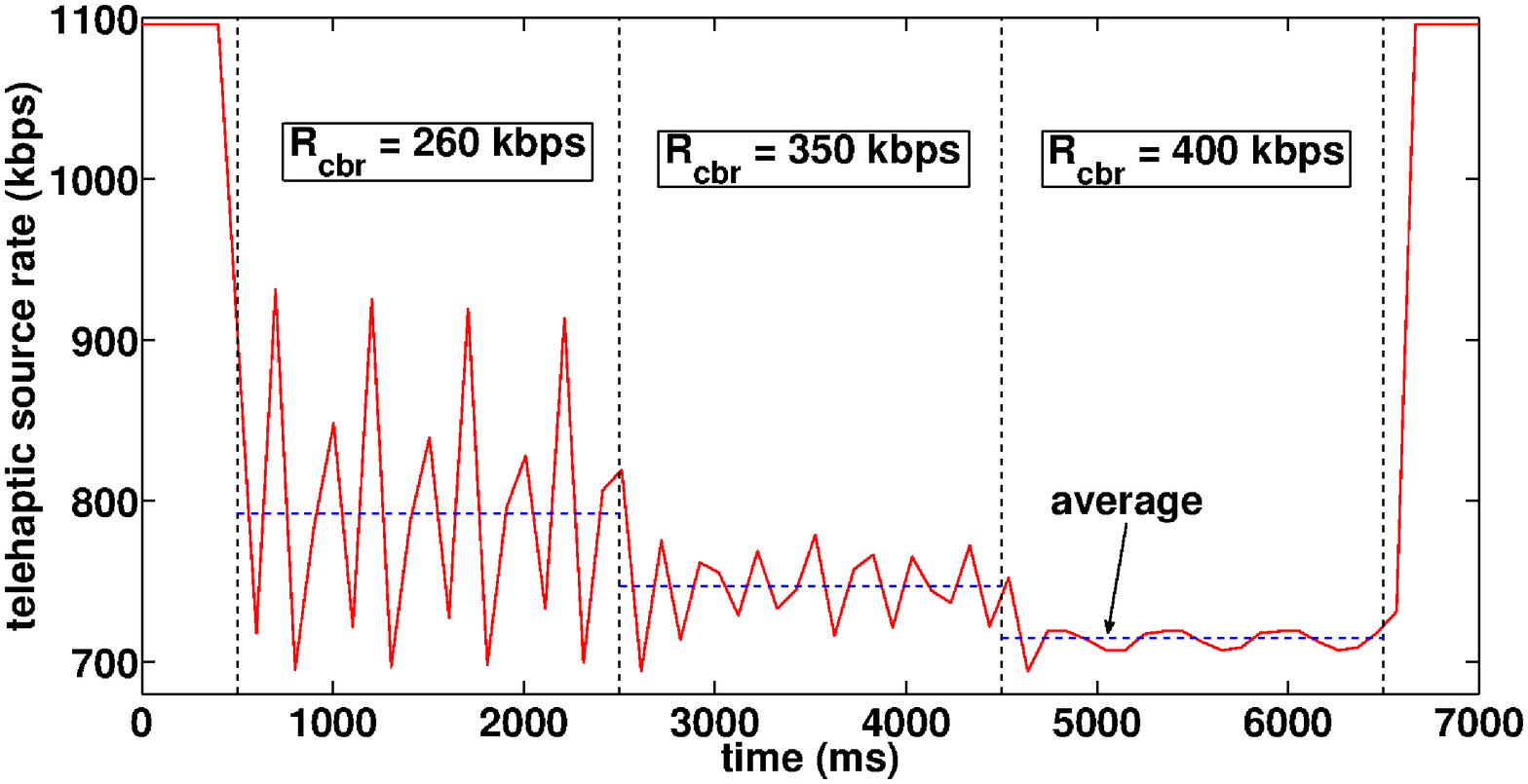}
\caption{Telehaptic source rate evolution under time-varying cross-traffic conditions.}
\label{fig:CBRpiecewise}
\end{center}
\end{figure}

\noindent \textbf{Adaptation to time-varying cross-traffic:} In order
to test the robustness of DPM to time-varying cross-traffic
conditions, we simulate three CBR sources on the backward channel:
$C_1$, $C_2$ and $C_3$ with data rates of 260 kbps, 90 kbps and 50
kbps, respectively. Each of these sources operates over a different
interval of time, resulting in an overall cross-traffic scheme shown
in Equation~(\ref{equ:cbr}).
\begin{equation}
\label{equ:cbr}
R_{cbr} =
   \begin{cases}
    0, & \text{for } \text{0} < t \leq \text{500 ms}\\
        \text{260} \text{ kbps }(C_1), & \text{for } \text{500 ms} < t \leq \text{2500 ms}\\
        \text{350} \text{ kbps }(C_1, \text{ and } C_2), & \text{for } \text{2500 ms} < t \leq \text{4500 ms} \\
        \text{400} \text{ kbps }(C_1, C_2 \text{ and } C_3), & \text{for } \text{4500 ms} < t \leq \text{6500 ms} \\
        \text{0}, & \text{for } t > \text{6500 ms}
   \end{cases} 
\end{equation}
Figure \ref{fig:CBRpiecewise} shows the
temporal variation of DPM source rate. Until 500 ms, DPM achieves its
peak rate since the network in uncongested. After 500 ms, the network
is unable to support the peak rate, and DPM automatically lowers the
telehaptic data rate to avoid congestion. Note that as $R_{cbr}$
increases, DPM lowers its transmission rate progressively. Once the
CBR cross-traffic is withdrawn at $t$ = 6500 ms, DPM reverts to its
peak rate. Thus, we see that DPM exhibits cross-traffic friendliness, and
performs a robust congestion control
under time-varying cross-traffic settings. \\
\begin{table}[h]
\centering
\begin{tabular}{|c|c|c|c|c|}%
      \hline
      & \multicolumn{2}{c|}{Max. Delay (ms)} & \multicolumn{2}{c|}{Max. Jitter (ms)}\\
      \hline
      & QoS & Observed & QoS & Observed\\
      \hline
      Haptic & 30 & 29.738 & 10 & 3.628 \\
      \hline
      Audio & 150 & 27.952 & 30 & 5.372 \\
      \hline
      Video & 400& 63.629 & 30 & 8.255 \\ 
      \hline
      
\end{tabular}
\caption{Comparison of the telehaptic delay and jitter observed for different media
for $R_{cbr}$ = 400 kbps, along with the corresponding QoS specifications.}
\label{table:delaysummary}
\end{table}

\noindent \textbf{Telehaptic delay and jitter measurements:}
Table~\ref{table:delaysummary} summarizes the observed telehaptic
delay and jitter for haptic, audio and video streams, respectively,
with $R_{cbr}$ = 400 kbps. It can be seen that even under heavy
cross-traffic conditions, DPM enables the telehaptic application to
comply with the QoS limits. 
Note that the measured haptic jitter is
3.628 ms (see Section \ref{subsubsec:networkbased} for an analysis),
which is significantly below the QoS jitter limit.\\

\noindent \textbf{Haptic signal reconstruction:} We now study the
effects of network cross-traffic, DPM and data extrapolation on the
haptic signal reconstruction at the OP. We compare the reconstructed
signal with that corresponding to an adaptively sampled strategy, and
measure the improvement in haptic signal display that DPM yields. For
this purpose, we use real telehaptic traces captured during the
telepottery experiment. Ten pilot telehaptic signals were used in the
evaluation of the proposed scheme, with each signal corresponding to a
different human subject. For brevity, we present results corresponding
to a particular pilot signal. We employ a Weber sampler with a
threshold of 12\% for adaptively sampling the force samples at the
teleoperator \cite{ref:psychophysics}. We use the standard zero-order
hold strategy for haptic data extrapolation. For this experiment, we
set $R_{cbr}$ = 400 kbps.

\begin{figure}[!h]
\begin{center}
\includegraphics[height = 40mm, width = 100mm]{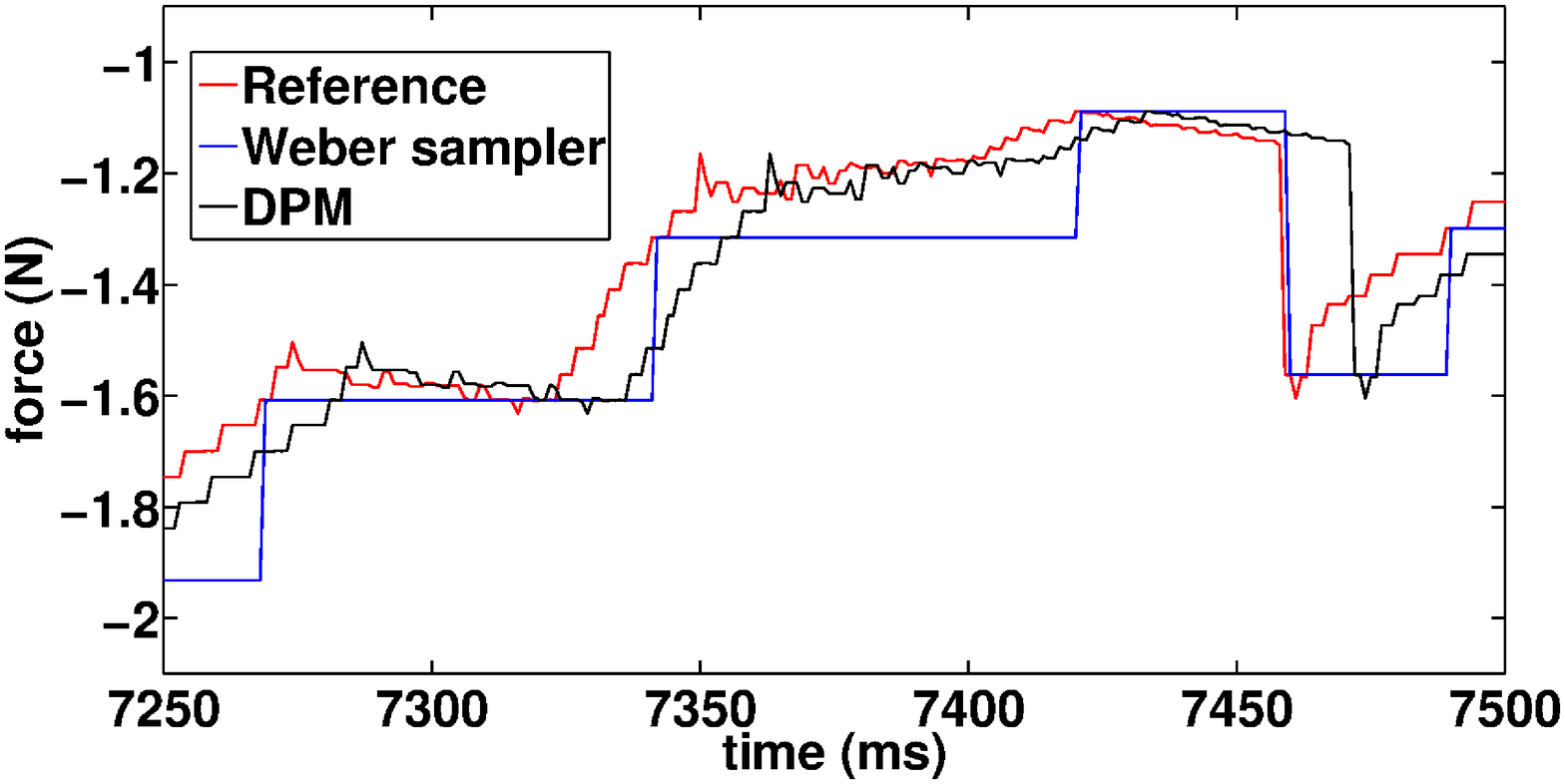}
\caption{Graph showing the reconstructed force signals at OP with Weber sampling and DPM for $R_{cbr} =$ 400 kbps.}
\label{fig:allinone}
\end{center}
\end{figure}

\begin{table}[!h]
\centering
\begin{tabular}{|c|c|c|}%
      \hline
       & \multicolumn{1}{c|}{SNR (dB)} & \multicolumn{1}{c|}{Improvement over WS (dB)} \\
      \hline
      Weber sampler (WS) & 21.5518 & -\\
      \hline
      DPM& 24.0986 &  2.5468 \\
      \hline
\end{tabular}
\vspace*{0.2 cm}
\caption{Comparison of SNR (in dB) in case of Weber sampler and DPM, with $R_{cbr} =$ 400 kbps on backward channel.}
\label{table:snrcomparison}
\end{table}

For benchmarking, we make use of a reconstructed signal captured using
an ideal (high bandwidth, zero jitter) network;
we treat this signal as the reference signal. Figure
\ref{fig:allinone} shows the force signal displayed at the OP under
different schemes. As expected, DPM, being a lossless protocol,
captures the fine details of the reference signal well. On the other
hand, the Weber sampled signal is a piecewise constant approximation
of the reference signal. It is to be noted that under the Weber
sampling strategy, `perceptually significant' samples are displayed
earlier at the OP as compared to DPM. This is because of the higher
packetization delay under DPM.
Using SNR as a performance metric to measure the reconstruction error
at the OP (against the reference signal),
Table~\ref{table:snrcomparison} compares the SNR (in dB) measured for
the reconstructed haptic signal under different schemes.
We see that DPM exhibits a substantial SNR improvement of around 2.5 dB over the
Weber sampling strategy. In our experiments, we have found a
comparable SNR improvement for other haptic traces.\\
 
\begin{figure}[!h]
\centering
  \subfloat[]{\includegraphics[height = 45mm, width = 70mm]{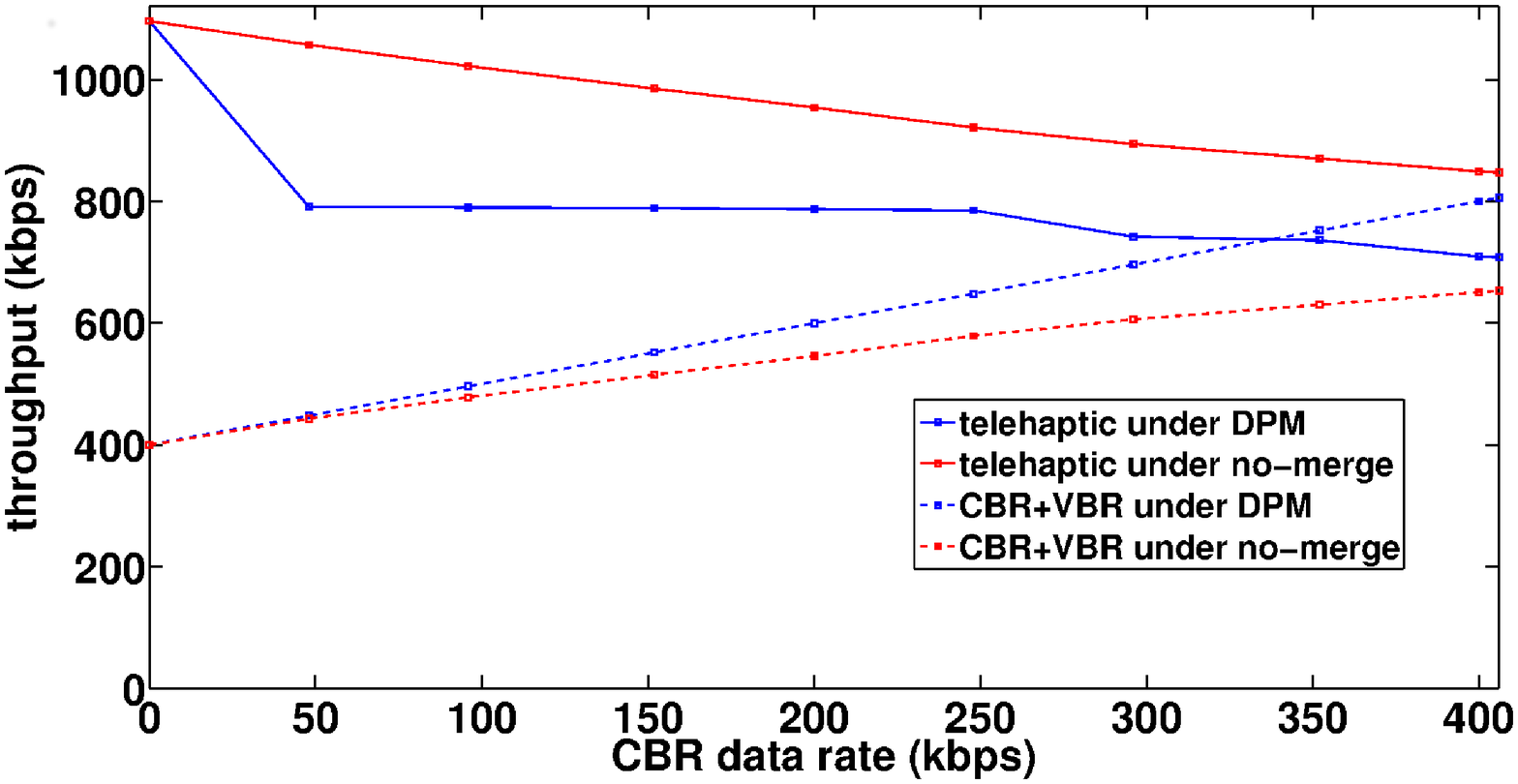}\label{fig:cbrthrput}}~ \hspace{-0.8cm}
  \subfloat[]{\includegraphics[height = 45mm, width = 70mm]{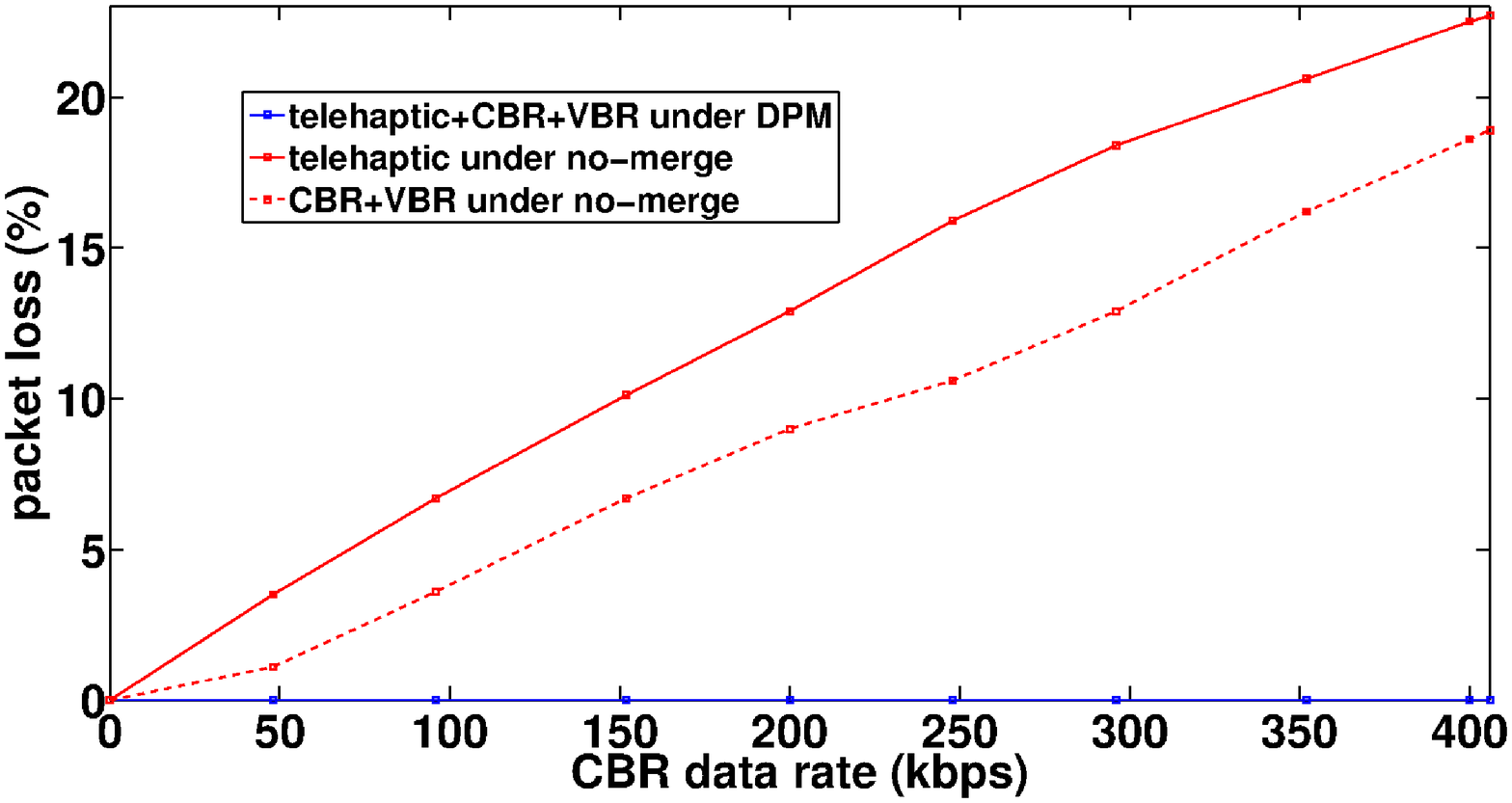} \label{fig:cbrloss}}
\caption{Telehaptic-CBR traffic interplay demonstrating the improvement of DPM over
no-merge in terms of (a) throughput and (b) packet loss.}
\label{fig:cbranalysis}
\end{figure}

%

\noindent \textbf{Throughput-Loss Measurements:} 
Figure \ref{fig:cbranalysis} compares the performance of DPM and the
no-merge scheme,\footnote{Recall that the no-merge scheme transmits at
  the peak telehaptic data rate, oblivious to the state of the
  network. In the literature, this scheme is also referred to as plain
  UDP \cite{ref:udp}.} in terms of throughput and packet losses, under
various CBR cross-traffic conditions. The results show that for
$R_{cbr} <$ 4 kbps, the two schemes exhibit similar behavior since the
network can sustain the peak telehaptic data rate. As $R_{cbr}$
increases further, the DPM appropriately lowers the telehaptic data
rate resulting in zero packet loss until $R_{cbr}$ approaches 406
kbps. On the other hand, the no-merge scheme demonstrates deteriorated
performance when $R_{cbr} >$ 4 kbps due to its network obliviousness. 

Figure \ref{fig:cbranalysis}(b) shows that the telehaptic and cross-traffic
streams sustain severe packet losses with increasing $R_{cbr}$ under
the no-merge scheme, whereas DPM avoids packet losses altogether by
adapting the telehaptic data rate to the intensity of cross-traffic.
We note that DPM is friendly to CBR and VBR cross-traffic. Indeed, the
cross-traffic streams see a higher throughput (and zero loss) under
DPM as compared to no-merge.\\

\noindent \textbf{DPM with hold-up:}
Motivated by Figure~\ref{fig:delayzones}(a), we propose a variant of
DPM that seeks to reduce the jitter induced by frequent rate
adaptations. Recall that in the experiment corresponding to
Figure~\ref{fig:delayzones}(a), the maximum data rate for the
telehaptic stream that would keep the bottleneck link stable
corresponds to $k = 2.$ However, when DPM experiences a steady delay
at $k = 2,$ it switches to $k = 1,$ which starts yet another cycle of
rate adaptations. In this case, it is clear that if DPM were to hold
on to the setting $k = 2$ for a longer period, there would be a
reduction in jitter at the receiver. This motivates the following
modification of DPM.

\emph{DPM with hold-up} is identical to DPM, except for the following
modification. It remembers the value of $k,$ say $\hat{k},$ that was
operating when the previous $I_c$ trigger was received. Subsequently,
once $k = \hat{k} + 1,$ the algorithm ignores $I_S$ triggers for a
\emph{hold-up duration} $T_h.$

Note that the hold-up modification would work well under steady or
slowly varying cross-traffic conditions. Indeed, if one assumes that
the cross-traffic is steady, then one may conclude that the previous
$I_C$ trigger was actually caused by the rate adaptation $\hat{k} + 1
\rightarrow \hat{k}.$ This suggests that $\hat{k} + 1$ is currently
the optimal operating point for the algorithm. Thus, once in this
state, \emph{DPM with hold-up} puts off attempts to increase its rate
further for a period $T_h.$ Of course, this modification is pessimal
in that it misses any opportunities for increasing the transmission
rate during the hold-up period $T_h.$

\vspace{-0.3cm}
\begin{figure}[!h]
\centering
  \subfloat[]{\includegraphics[height = 40mm, width = 75mm]{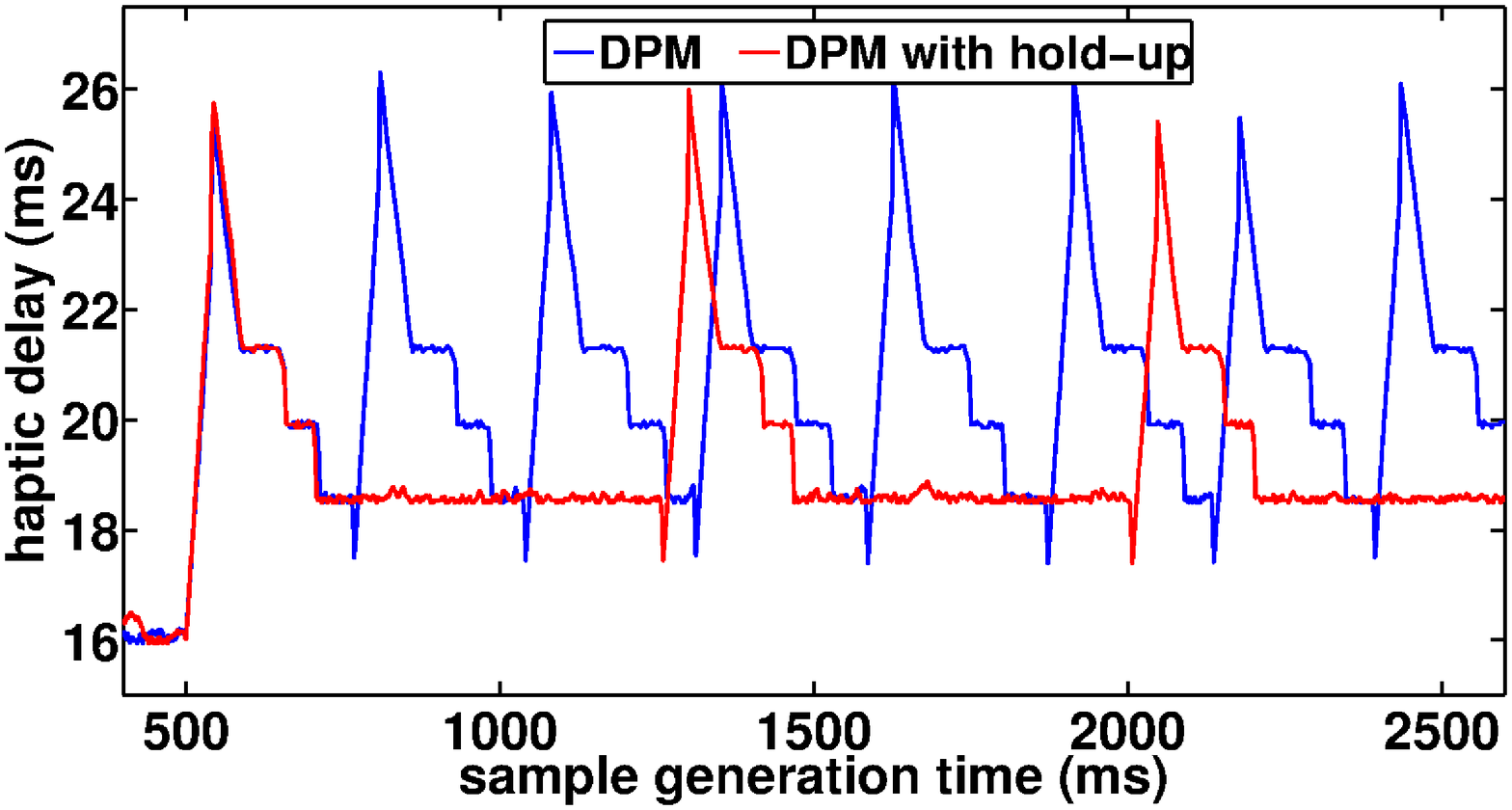} \label{fig:holdnohold2}}~ \hspace{-1cm}  
  \subfloat[]{\includegraphics[height = 40mm, width = 75mm]{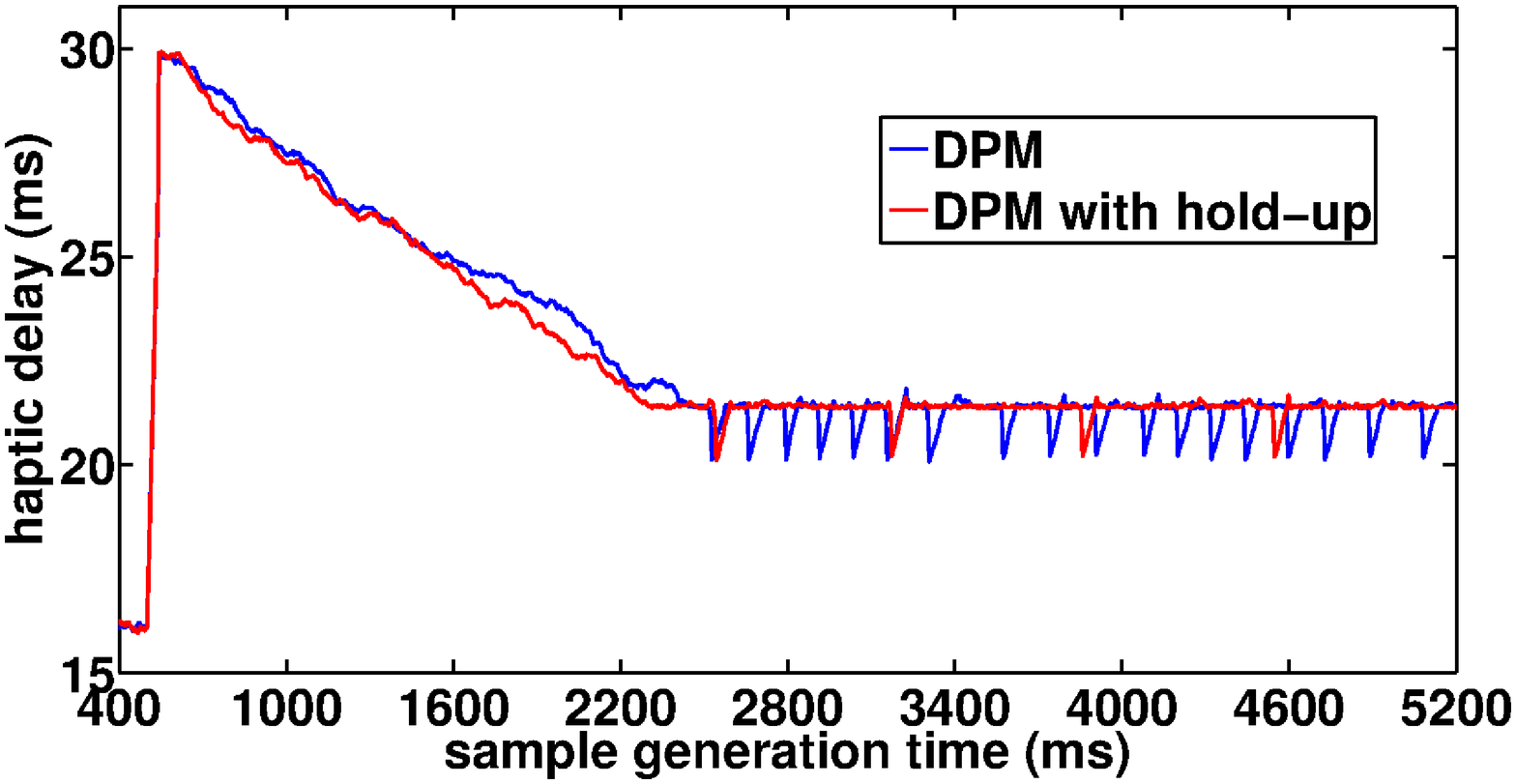} \label{fig:holdnohold4}}
\caption{Haptic delay plots for DPM and DPM with hold-up techniques in presence of CBR cross-traffic (a) 260 kbps (b) 400 kbps.}
\label{fig:holdVSnohold}
\end{figure}
\vspace{-0.3cm}
Figure \ref{fig:holdVSnohold} shows the haptic delay variation plots
for DPM with and without hold-up in case of $R_{cbr} =$ 260 kbps and
400 kbps. $T_h$ is heuristically chosen to be 500 ms.  As expected,
under the hold-up modification, the cycles of delay fluctuation occur
less frequently.
The average jitter for DPM and DPM with hold-up for $R_{cbr} =$ 400
kbps are measured as 1.3 $\mu$s and 0.93 $\mu$s, respectively. This
implies a reduction in average jitter of around 29\% over DPM. The SNR
of the reconstructed signal under DPM with hold-up is measured to be
24.8332 dB, which is around 0.7 dB higher than the SNR under DPM (see
Table~\ref{table:snrcomparison}).

In conclusion, when it is known a priori that the cross-traffic is
slowly varying, the hold-up modification provides a modest QoS
improvement over DPM.

\subsection{Telepottery Subjective Grading}
\label{subsec:percepexptresults}
\begin{figure}[!h]
\begin{center}
\includegraphics[height = 40mm, width = 110mm]{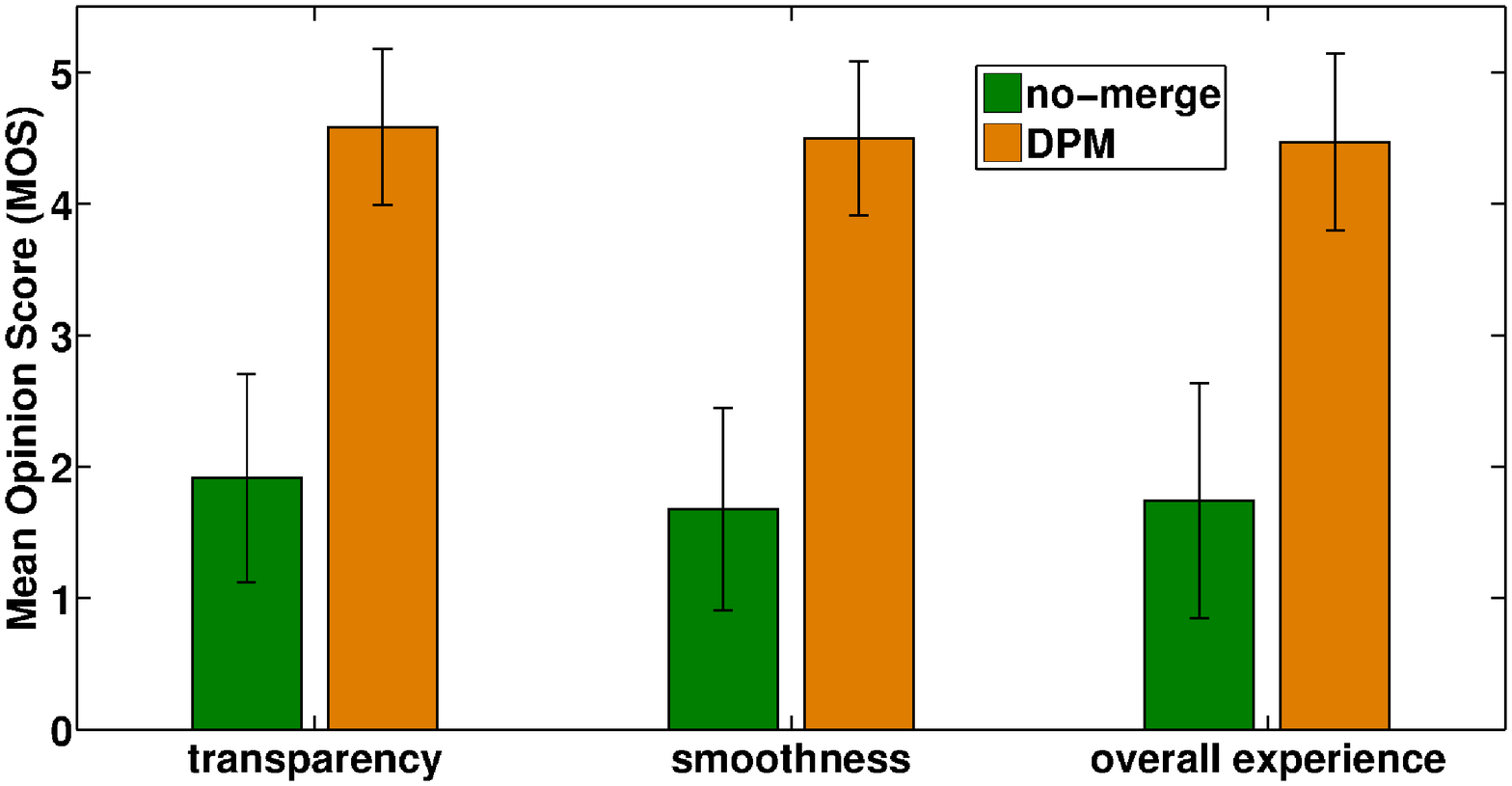}
\caption{MOS of the subjective evaluation of the proposed technique on three specific perceptual parameters, averaged over twenty human subjects.
The vertical bars denote the standard deviation of the subject grades.}
\label{fig:DCRgrading}
\end{center}
\end{figure}
We now move to the qualitative results of the real-time telepottery task.
Figure \ref{fig:DCRgrading} presents the mean opinion score (MOS) of
the DCR recorded with twenty human subjects for each of the three
perceptual parameters i.e., transparency, smoothness and overall
experience. We observe that the MOS recorded while using the no-merge
technique is less than 2, which corresponds to an annoying user
experience. In fact, a few subjects found the no-merge experience so
disturbing that they hardly made any contact with the clay model.
When DPM is employed, the MOS under each of the three perceptual
parameters improves substantially (in the neighborhood of 4.5,
signifying nearly imperceptible degradation in the user
experience). 

In order to statistically evaluate the improvement in the perception of DPM
over the no-merge scheme, we perform paired $t$-test over the subject grades.
The test results corresponding to the aforementioned perceptual parameters are as follows:
(i) transparency - $t(19) = 10.81, p < 0.001$; (ii) smoothness - $t(19) = 13.97, p < 0.001$;
and (iii) overall experience - $t(19) = 11.72, p < 0.001$.
This further substantiates our claim that the rate adaptation
mechanism of DPM introduces negligible perceivable artifacts.

Thus, we conclude that DPM preserves the immersiveness of the
telepottery activity in spite of heavy cross-traffic on the network,
thereby resulting in a user-friendly and enjoyable telepottery
experience.

\ignore{This implies that the participants felt the experience with
  the DPM scheme was of identical/acceptable standards, whereas the
  no-merge scheme resulted in a highly degraded perceptual quality.

  Hence DPM achieves significant improvement in the human perception
  of telehaptic objects compared to the network-oblivious no-merge
  scheme.}

\subsection{Comparison with existing telehaptic communication
  techniques}
\label{sec:comparison}

In this section, we compare the performance of DPM with RTP \cite{ref:rtp}
and other recently proposed telehaptic communication protocols.
\vspace{-0.4cm}
\begin{figure}[!h]
\begin{minipage}[b]{0.45\linewidth}
\centering
\includegraphics[height = 40mm, width = 65mm]{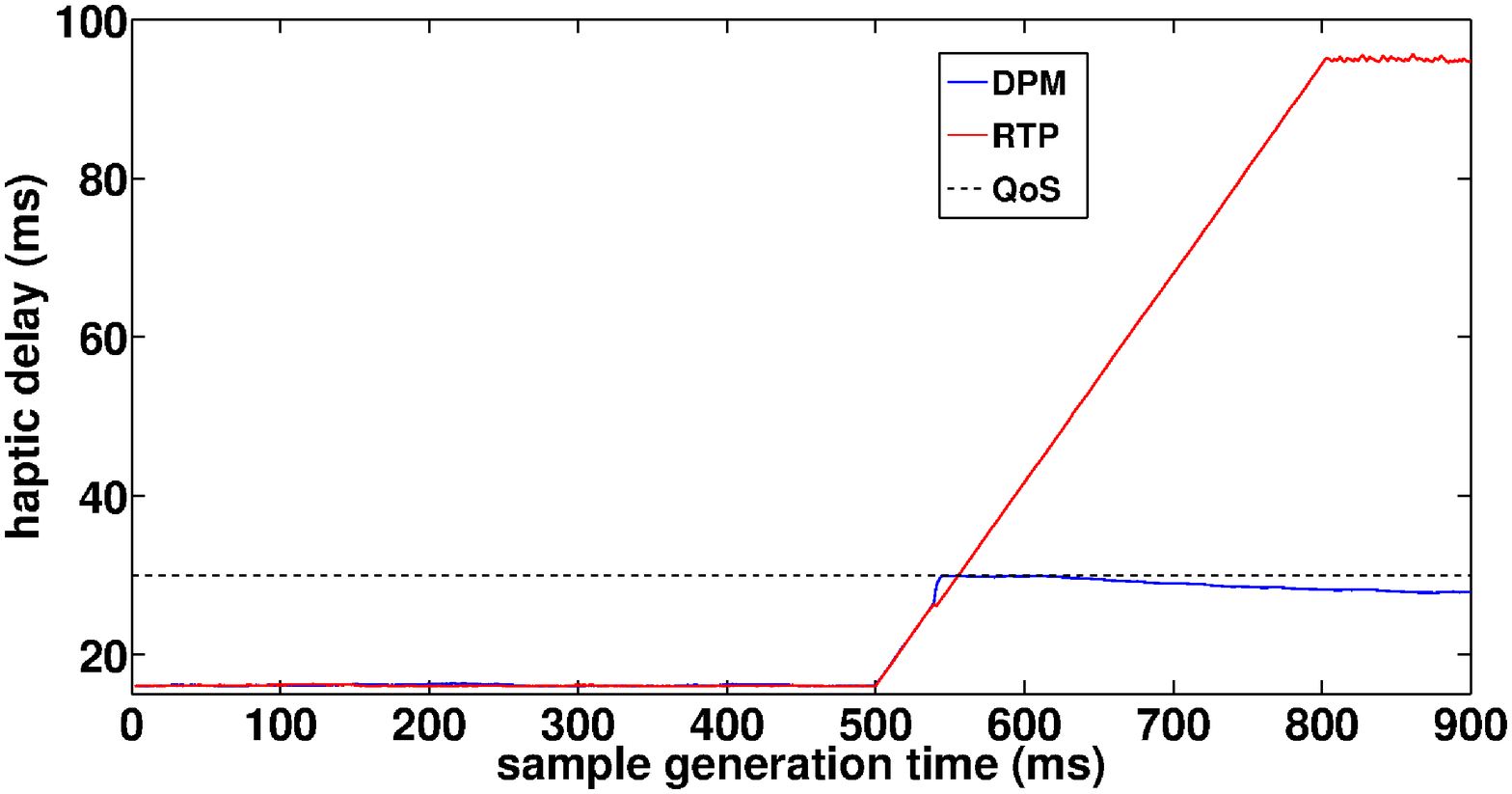}
\caption{Early congestion detection and responsiveness of DPM as opposed to sluggish behavior of RTP.}
\label{fig:dpmVSrtp}
\end{minipage}
\hspace{0.25cm}
\begin{minipage}[b]{0.45\linewidth}
\centering
\includegraphics[height = 40mm, width = 65mm]{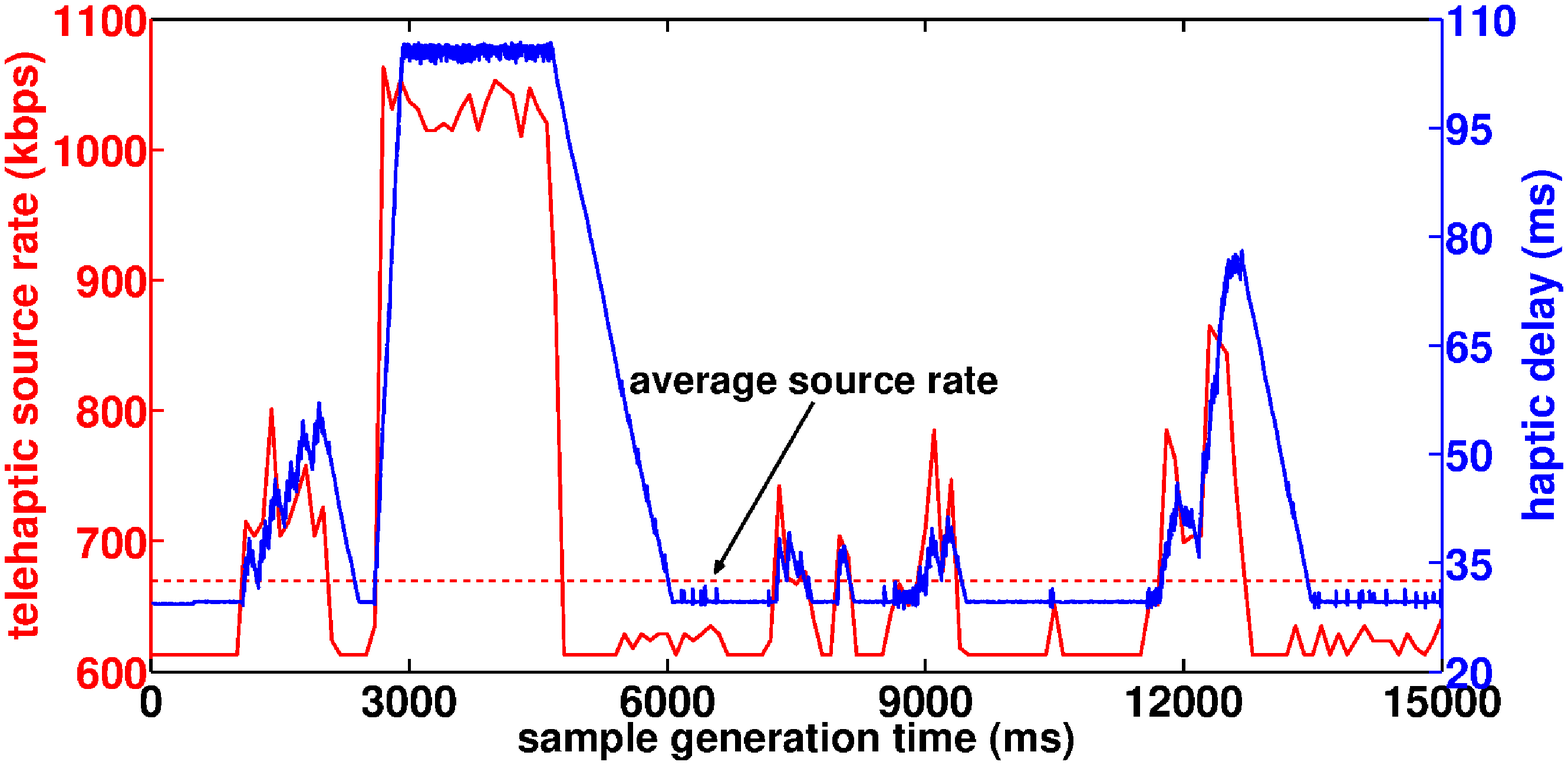} 
\caption{Telehaptic rate-delay plot of visual-haptic multiplexing on backward channel.}
\label{fig:vhmux}
\end{minipage}
\end{figure}
\begin{figure}[t]
\begin{center}
\end{center}
\end{figure}
\vspace{-0.3cm}

\noindent \textbf{1) Real-time transport protocol (RTP)}: We begin by
comparing DPM with RTP, which is the predominant protocol for media
streaming applications on the internet. We use the simulation setup
from Section~\ref{subsec:simresults}, with $R_{cbr} =$ 400 kbps on the
backward channel. Figure~\ref{fig:dpmVSrtp} shows the variation of the
end-to-end delay experienced by haptic samples with the sample
generation time. Note that once the CBR cross-traffic is introduced at
500 ms, DPM performs a prompt rate adaptation, maintaining end-to-end
delays below the QoS deadline of 30 ms. In the same setting, RTP
generates its first and the second RTCP reports at 500 ms and 1000 ms,
respectively. Since any rate-control mechanism based on RTP would not
make a rate-adaptation prior to 1000 ms, the haptic delays under any
such protocol would keep growing as shown in
Figure~\ref{fig:dpmVSrtp}, violating the QoS deadlines. Note that
network queues build up on the timescale of tens of
milliseconds. Thus, for telehaptic applications, RTP, which generates
network feedback reports every 500 ms, is too slow to allow for timely
rate-adaptation.\\

\vspace{-0.1cm}
\noindent \textbf{2) Visual-haptic multiplexing}: We now evaluate DPM
against the visual-haptic multiplexing scheme \cite{ref:cizmeci},
which employs the Weber sampler for force updates on the backward
channel. For this evaluation, we use the simulation setup from
Section~\ref{subsec:simresults} with $R_{cbr} =$ 400
kbps. Figure~\ref{fig:vhmux} shows the source rate evolution and
the resulting haptic delays under visual-haptic multiplexing obtained using one of the
traces from our real-time telepottery experiment. We see that even
though the available capacity on the backward channel (700 kbps) exceeds the
average transmission rate on the backward channel (670 kbps), the instantaneous rate
fluctuates substantially, resulting in occasional
QoS violations; for example, see the interval from 3000 to 6000
ms. Indeed, during times when the operator's movements are fast,
almost every force sample becomes perceptually significant, causing a
Weber sampler's instantaneous transmission rate to far exceed its
time-average. It is also worth noting that packet loss measured
between 3000 ms and 6000 ms is around 16\%, which could potentially
lead to significant perceptual degradation. In contrast, under the
same network conditions, the results of
Section~\ref{subsec:simresults} show that DPM meets the QoS
constraints and results in zero packet loss.

\ignore{Figure \ref{fig:vhmux} demonstrates that especially when the
  network is provisioned for the average source rate.  It can be
  observed that the source rate is in the neighborhood of peak rate
  when the operator's movements are fast, causing every force sample
  to be perceptually significant. On the other hand when the force
  signal is slowly varying, the sampler discards a large proportion of
  force updates leaving only the audio and video frames to be
  transmitted.  }
  \vspace{-0.5cm}
  \begin{figure}[!h]
\centering
  \subfloat[]{\includegraphics[height = 40mm, width = 70mm]{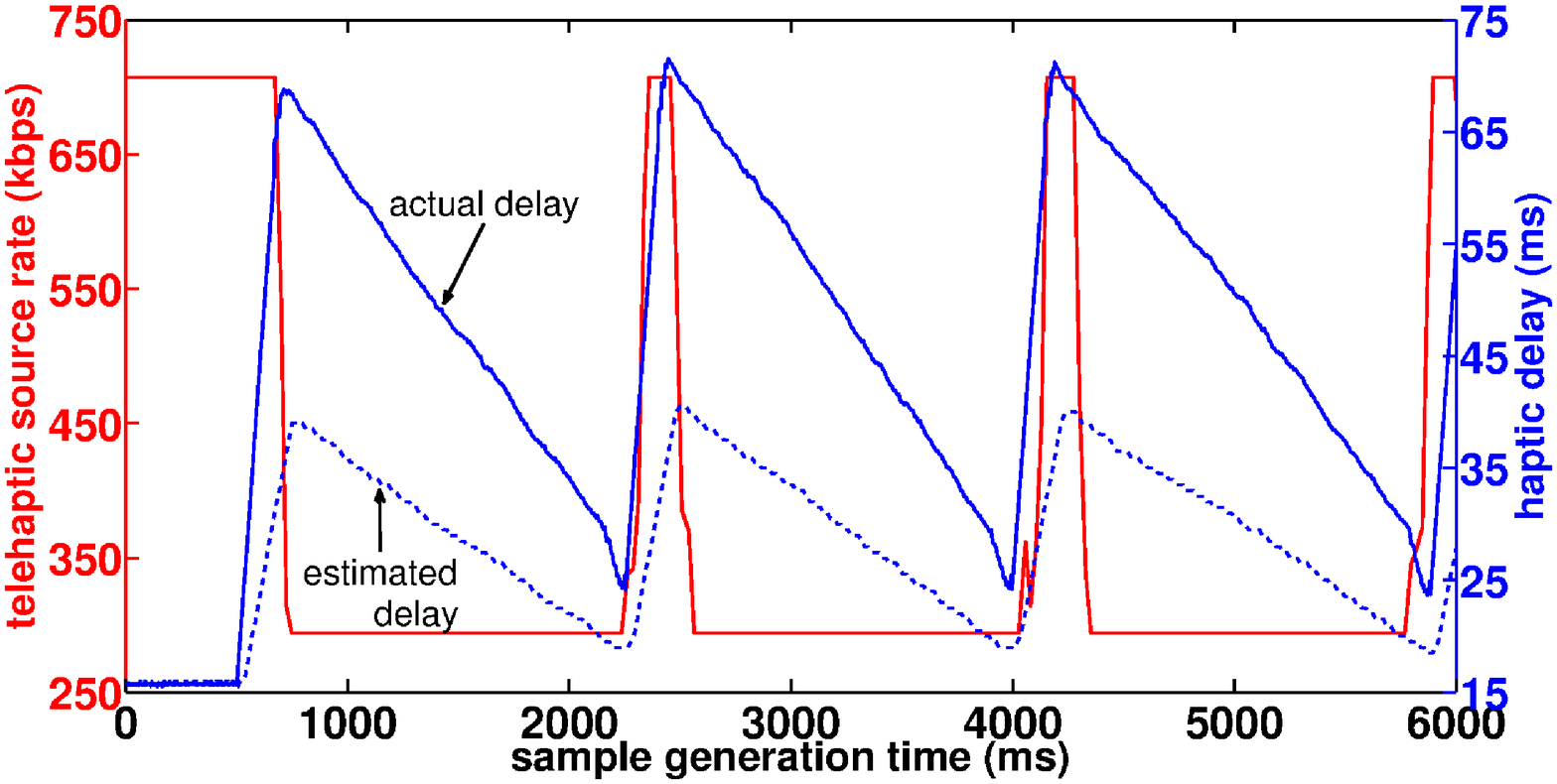}\label{fig:nafcah}}~\hspace{-0.4cm}
  \subfloat[]{\includegraphics[height = 40mm, width = 70mm]{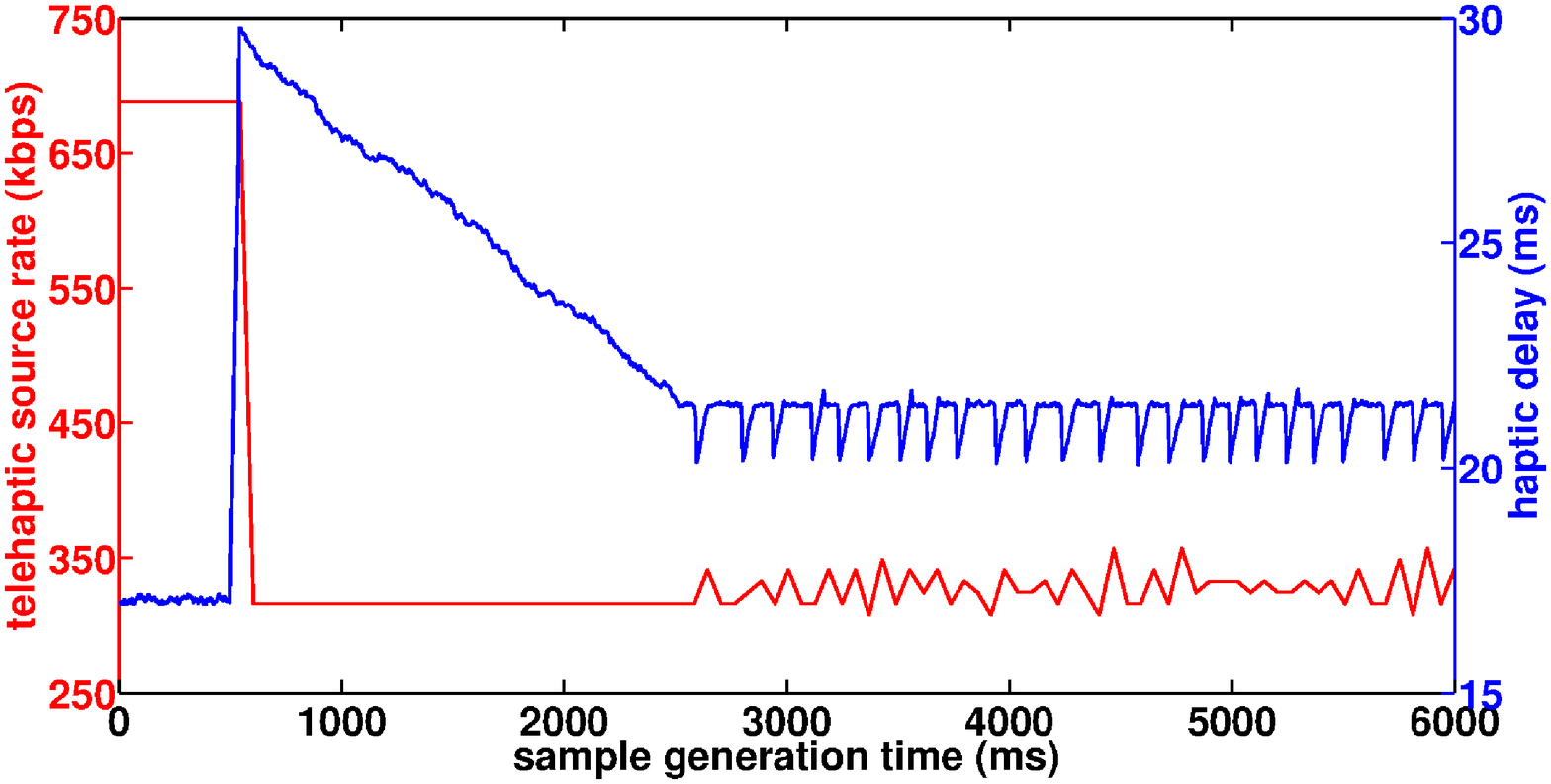} \label{fig:dpmfwd}}
\caption{Telehaptic rate-delay plots for the forward channel with (a) NAFCAH (b) DPM.}
\label{fig:dpmvsnafcah}
\end{figure}
  \vspace{-0.4cm}
\noindent \textbf{3) Network adaptive flow control algorithm for
  haptic data (NAFCAH)}: We now compare the performance of NAFCAH
\cite{ref:nafcahkokkonis}, a protocol that performs RTT-based rate
adaptation on the \emph{forward channel}, with DPM. We use the
simulation setup from Section~\ref{subsec:simresults}, except that the
CBR cross-traffic intensity on the forward channel is increased to 780
kbps; this makes the forward channel highly congested. With the
probing packet frequency of NAFCAH set as 100~Hz,
Figure~\ref{fig:nafcah} shows the evolution of the source transmission
rate and the delay experienced by the haptic samples under
NAFCAH. Note that NAFCAH incurs substantial QoS violations. The
reasons for this are twofold: Firstly, once congestion is detected,
NAFCAH cuts its transmission rate in stages (i.e., it employs a
\emph{multistep-increase} approach). As discussed in
Section~\ref{subsec:simresults}, this results in a relatively sluggish
congestion control. Secondly, NAFCAH uses round-trip-time measurements
to estimate congestion on the forward channel. This leads to incorrect
delay estimations under asymmetric network conditions, as shown in
Figure~\ref{fig:dpmvsnafcah}(a).

In contrast, as seen in Figure~\ref{fig:dpmvsnafcah}(b), DPM satisfies the QoS
constraints well under the same network conditions, thanks to its
aggressive \emph{step-increase} mechanism for rate reduction, and its
accurate end-to-end delay estimation mechanism.

\ignore{Since the forward channel telehaptic payload is significantly
  lower compared to that on the backward channel, the CBR intensity on
  the forward channel has to be higher in order to cause
  congestion. We note that under the current network setting, DPM is
  effective on the forward channel for $R_{cbr} \in$ [412, 784]
  kbps. For this experiment, we take $R_{cbr} = 780$ kbps on the
  forward channel and probing packet frequency of NAFCAH as 100
  Hz. NAFCAH follows multistep rate reduction resulting in a larger
  end-to-end delay before congestion control as demonstrated in
  \ref{fig:nafcah}.  In addition, it can be observed that the
  RTT-based network feedback underestimates the current network
  delay. This is true with any RTT-based estimation scheme in an
  asymmetric network setting.  Therefore, NAFCAH prematurely increases
  the data rate even before the buffers are flushed completely,
  leading to further congestion and frequent QoS violations. On the
  other hand, DPM performs precise network estimation and swift rate
  adaptation leading to delay QoS adherence as shown in Figure
  \ref{fig:dpmfwd}.}

\section{Conclusions and Limitations}
\label{sec:conclusions}
In this paper, we presented DPM, a transport layer congestion control
protocol for a lossless, real-time telehaptic communication.
In order to enable DPM to quickly respond to network variations, we
proposed the network feedback module for communicating the the end-to-end delays
to the transmitters with negligible overhead.
Via extensive simulations, we showed that DPM meets the QoS
requirements of telehaptic applications even under highly congested
network conditions. We also validated DPM's ability to provide a
seamless and immersive user experience over a congested network via a
real-time telepottery experiment with human subjects. Finally, we
showed that DPM outperforms previously proposed telehaptic
communication protocols.


  While the present paper explores the interplay between DPM
  and network-oblivious UDP traffic, the interplay between DPM and
  other network-aware cross-traffic streams (predominantly TCP)
  remains unexplored.  Further, it is not clear as to how multiple DPM
  streams coexisting on a network would share the available
  bandwidth. Finally, the implications of SNR improvement of our
  scheme over Weber sampling on the quality of the telehaptic task has
  not been investigated. We would like to address these issues in a
  future extension of this paper.



\bibliographystyle{ACM-Reference-Format-Journals}
\bibliography{acmref}




\appendix
\section*{APPENDIX}

\section{Application Layer Frame Structure}
\label{sec:applicationheader}

While the proposed protocol performs a transport layer
  function, in our implementation, we code the protocol in the
  application layer leveraging UDP at the transport layer. In this
  section, we describe the various application layer header fields of
  a telehaptic packet in our implementation. Table
\ref{table:packetstructure} shows the proposed application layer frame
structure for telehaptic communication. The topmost row is shown for
convenience, to indicate the bit positions in the frame.  The frame
structure starts with the field \textit{M}. The haptic header size is
8 bytes, whereas the audio and video headers consume 5 bytes each.
Since the focus is only on augmenting either audio or video with
haptic data, the effective application layer header size is 13 bytes.
The audio and video related headers are included only in presence of
audio and video payload, respectively.  Table
\ref{table:fielddescribe} describes each of the header fields in
detail.  The telehaptic payload includes haptic-audio/video payload
based on the value indicated in the field $M$. Haptic payload on the
forward channel includes position and velocity information of the
operator, whereas the backward channel carries force information.

\begin{table*}[htbp]
\begin{center}
\resizebox{12cm}{1.2cm}{
\begin{tabular}{*{32}{|c}}
      \hline
      0 & 1 & 2 & 3& 4& 5 & 6 & 7& 8&9 & 0 & 1 & 2 & 3& 4& 5 & 6 & 7& 8&9 & 0 & 1 & 2 & 3& 4& 5 & 6 & 7& 8&9 & 0 &\multicolumn{1}{c|}{1} \\ \hline
      \multicolumn{3}{|c}{M} & \multicolumn{3}{|c|}{k} & D & X & \multicolumn{24}{c|}{Notification Delay}\\
      \hline
      \multicolumn{32}{|c|}{Haptic Sample Timestamp}\\
      \hline
      \multicolumn{16}{|c}{Audio Frame No.} & \multicolumn{16}{|c|}{Audio Payload Size}\\
      \hline
      \multicolumn{16}{|c}{Video Frame No.} & \multicolumn{16}{|c|}{Video Payload Size} \\
      \hline
      \multicolumn{8}{|c|}{Audio Fragment No.} & \multicolumn{8}{c}{Video Fragment No.} & \multicolumn{16}{|c|}{Telehaptic Payload}\\
      \hline
      \multicolumn{32}{|c|}{Telehaptic Payload}\\
      \multicolumn{32}{|c|}{..........}\\
      \hline
\end{tabular}
}

\end{center}
\caption{Telehaptic packet format at the application layer. The top row is numbered bitwise for illustration. }
\label{table:packetstructure}
\end{table*}

\begin{table}[htbp]
\begin{center}
\resizebox{\columnwidth}{5cm}{
     \begin{tabular}{ | p{2.5cm} | p{1cm} |  p{10cm} |}
    \hline
    \centering Field & Bits & Description \\ \hline
    \centering$M$ & 3 & Indicates the type of media data contained in the payload. 0: haptic, 1: haptic-audio, 2: haptic-video. The additional bit is included to provide support for additional media types.\\ \hline
    \centering$k$ & 3 & Indicates the $k$-merge scheme used for the current packet. \\ \hline
    \centering $D$ & 1 & Delay indicator field. Indicates the transmission status of the delay embedded in the packet header. 0 - fresh transmission, 1 - repetitive transmission. \\ \hline
    \centering X & 1 & Reserved for future enhancements to the protocol. \\ \hline
    \centering Notification Delay & 24 & End-to-end delay inserted by the in-header delay notification mechanism. \\ \hline
    \centering Haptic Sample Timestamp & 32 & Indicates the generation time (in ms) of the haptic sample in the payload in case of $k$ = 1. In case of a higher $k$, this field indicates generation time of the earliest of the $k$ haptic samples in the payload. \\ \hline         
    \centering Audio Frame No. \& Audio Fragment No. & 16 \& 8 & Indicates frame number of the current audio payload and fragment number of the current audio frame, respectively. \\ \hline
    \centering Video Frame No \& Video Fragment No. & 16 \& 8 & Indicates frame number of the current video payload and fragment number of the current video frame, respectively. \\ \hline
    \centering Audio \& Video Payload Size & 16 \& 16 & Indicates the size of the audio and video payload in bytes, respectively.\\ \hline
    \end{tabular}
}
\end{center}
\caption{Detailed description of the application layer header structure.}
\label{table:fielddescribe}
\end{table}

\section{Characterization of maximum haptic delay}
\label{sec:upperbound}

In this section we derive an expression for the maximum end-to-end
delay experienced by haptic samples under DPM over a single bottleneck
network topology (see Figure~\ref{fig:topology}) with CBR cross
traffic.\footnote{Note that since our protocol operates at the
  transport layer (TL), we characterize the maxiumum TL-TL latency,
  i.e., the maximum latency between the arrival of a haptic sample at
  the TL of the sender and the reception of the sample at the TL of
  the receiver.} This analytical characterization enables us to
identify the class of network configurations where QoS-compliant
telehaptic communication is feasible.

Let $R_{cbr}$ denote the CBR cross-traffic intensity (in kbps) on the
channel under consideration. For simplicity, we assume that the
reverse channel is uncongested, so that the packetization rate on the
reverse channel equals~1~kHz. Recall that $\tau$ and $\mu$ denote the
one way propagation delay (in ms) and the bottleneck channel capacity
(in kbps), respectively. Define $$k_{opt} = \min\{k \in
\{1,2,\cdots,k_{max}\}: R_k + R_{cbr} \leq \mu\}.$$ Note that when DPM
operates at $k \geq k_{opt},$ the bottleneck link remains
uncongested. It then follows that in steady state, the maximum
end-to-end delay is experienced during the buffer build-up that
results from DPM switching from $k = k_{opt}$ to $k = k_{opt}
-1.$

For simplicity, let us denote the instant when DPM sets $k =
k_{opt} -1$ by $t = 0.$ Let $d_{inc}$ denote the generation time of
the resulting congestion trigger. 
Note that $d_{inc}$ is the time
required for the delay measurement corresponding to the $N$th packet
transmitted after $t = 0$ to arrive at the transmitter. We can write
an expression for $d_{inc}$ as follows.
\begin{equation*}
  d_{inc} = N(k_{opt}-1)(1\mathrm{ms}) + \frac{N(R_{cbr}+R_{k_{opt}-1} -\mu)}{\mu}(k_{opt}-1)(1\mathrm{ms})+ 2\tau  + 1\mathrm{ms}
 \label{equ:dinc}
\end{equation*}
The first term above is the generation time of the $N$th packet after
time~0. The second term is the queueing delay seen by this packet. The
third term equals the round trip propagation delay, and the fourth term
($1\mathrm{ms}$) represents the time gap between arrival of the $N$\textsuperscript{th}
packet at the receiver and the piggybacking of its delay on the
reverse channel.

Since the queue at the ingress of the bottleneck link builds up at the
rate of $R_{cbr}+R_{k_{opt}-1}-\mu$ until time $d_{inc}$, we obtain
the following expression for the maximum queue occupancy.
\begin{equation*}
  q_{inc}  = (R_{cbr}+R_{k_{opt}-1}-\mu)d_{inc}
 \label{equ:qinc}
\end{equation*}
The maximum end-to-end delay would be clearly experienced by the
packet that sees a queue occupancy of $q_{inc}$. This leads us to the
following expression for the maximum end-to-end delay:
\begin{equation*}
  d_{hap} = \tau+\frac{q_{inc}}{\mu}+(k_{opt}-1)(1\mathrm{ms})
\end{equation*}
The first term above captures the one-way propagation delay, the
second term captures the maximum queueing delay, and the last term captures
the packetization delay seen by the earliest haptic sample in the
packet. Combining the above equations, we get:
\begin{equation}
  \label{eq:max_haptic_delay}
\begin{array}{rl}
  d_{hap} &= \tau + (k_{opt}-1)(1\mathrm{ms}) + \left[N(k_{opt}-1)(1\mathrm{ms}) + 2\tau + 1\mathrm{ms}\right]\left(\frac{R_{cbr}+R_{k_{opt}-1} - \mu}{\mu}\right) \\
  &\quad \quad + N(k_{opt}-1)(1\mathrm{ms}) \left(\frac{R_{cbr}+R_{k_{opt}-1} - \mu}{\mu}\right)^2
\end{array}
\end{equation}
We have validated the accuracy of the above expression via
simulations. 

Note that Equation~\eqref{eq:max_haptic_delay} enables us to
characterize the set of link capacities, propagation delays, and
cross-traffic intensities that satisfy the haptic QoS constraints. In
the following section, we relate the maximum end-to-end haptic delay
to the maximum end-to-end delay seen by the audio and video streams.

\section{Characterization of the maximum audio/video delay}
\label{sec:AVbound}

In this section, we derive an upper bound on the maximum end-to-end
audio/video delays under DPM over a single bottleneck network topology
(see Figure~\ref{fig:topology}) with CBR cross-traffic.\footnote{Note again that
  we only consider the maximum TL-TL latency.} Interestingly, these
upper bounds involve the maximum haptic delay $d_{hap}$ characterized
in Appendix~\ref{sec:upperbound}. Thus, we are able to relate haptic QoS
compliance to QoS compliance for audio and video.

Recall that $f_a$ and $f_v$ represent the (peak) frame rates (in
frames per second) of audio and video, respectively. Also, $s_a$ and
$s_v$ represent the (peak) audio and video frame sizes (in bytes),
respectively. Finally, $s_m$ represents the size of the audio/video
fragment (in bytes) in each telehaptic fragment (see
Section~\ref{subsec:mediamultiplex}).

Note that our media multiplexing framework guarantees that at the
instant an audio frame is generated, the previous audio frame has
already been multiplexed with the haptic stream. Thus, the
multiplexing latency seen by the audio frame equals
$\frac{s_a}{s_m}(1\mathrm{ms}).$ There is an additional packetization
latency that is at most $(k_{max}-1)(1\mathrm{ms}).$ Finally, the
maximum end-to-end delay experienced by the packet is equals
$d_{hap}.$ This yields the following upper bound on the (TL-TL) audio
delay.
\begin{equation}
  d_{aud} \leq d_{hap} + \frac{s_a}{s_m}(1\mathrm{ms})+ (k_{max}-1)(1\mathrm{ms})
 \label{equ:aDelay}
\end{equation}

Next, we move to the maximum delay experienced by a video frame
(TL-TL). Our multiplexing framework guarantees that by the time a
video frame is generated, the previous one has been multiplexed. Thus,
the maximum multiplexing delay equals $\frac{1}{f_v}.$ Adding to this
the maximum packetization delay and the maximum end-to-end delay
experienced by a packet, we obtain the following upper bound on the
TL-TL video delay.
\begin{equation}
  d_{vid} \leq d_{hap} + \frac{1}{f_v} + (k_{max}-1)(1\mathrm{ms})
 \label{equ:vDelay}
\end{equation}

From Equations (\ref{equ:aDelay}) and (\ref{equ:vDelay}), we can
compute the maximum delay seen by audio/video frames assuming that the
QoS constraint on haptic delay is satisfied, i.e., $d_{hap} \leq$
30 ms. Consider the settings assumed in our simulations: $f_a = 50
\text{ fps }, f_v = 25 \text{ fps }, s_a = 160 \text{ B }, s_v = 2
\text{ kB }.$ This leads to $s_m = 58 \text{ B }$. It then follows
from Equations~(\ref{equ:aDelay}) and~(\ref{equ:vDelay}) that $d_{aud}
\leq $ 35.75 ms, $d_{vid} \leq$ 73 ms. Note that these bounds are well
below the audio/video QoS targets. Thus, under the proposed protocol,
meeting the (strict) haptic delay constraint in general leads to
compliance with the audio/video delay constraint.

\end{document}